\documentclass[prb,aps,amsmath,amssymb,amsfonts,floatfix,superscriptaddress,
showpacs,twocolumn]{revtex4}
\usepackage{graphicx}
\usepackage[dvips]{color}
\usepackage{dcolumn}
\newcommand{\bw}[1]{\raisebox{1.5ex}[-1.5ex]{#1}}

\newcommand{\bk}{\mathbf{k}}
\newcommand{\bR}{\mathbf{R}}
\newcommand{\iomn}{i\omega_n}

\def\bavs3{BaVS$_3$}
\def\Tmit{$T_{\rm MIT}\,$}
\def\t2g{$t_{2g}$}
\def\A1g{$A_{1g}$}

\begin{document}
\title{Competing itinerant and localized states in strongly correlated BaVS$_3$}
\author{Frank Lechermann}
\affiliation{I. Institut f{\"u}r Theoretische Physik, Universit{\"a}t Hamburg, 
D-20355 Hamburg, Germany}
\author{Silke Biermann}
\affiliation{CPHT, {\'E}cole Polytechnique, 91128 Palaiseau Cedex, France}
\author{Antoine Georges}
\affiliation{CPHT, {\'E}cole Polytechnique, 91128 Palaiseau Cedex, France}

\begin{abstract}
The electronic structure of the quasi-lowdimensional vanadium sulfide \bavs3
is investigated for the different phases above the magnetic ordering 
temperature. By means of density functional theory and its combination with
dynamical-mean field theory, we follow the evolution of the relevant low-energy 
electronic states on cooling. Hence we go in the metallic regime from the room 
temperature hexagonal phase to the orthorhombic phase after the first structural 
transition, and close with the monoclinic insulating phase below the 
metal-insulator transition. Due to the low symmetry and expected intersite 
correlations, the latter phase is treated within cellular dynamical mean-field 
theory. It is generally discussed how the intriguing interplay between 
band-structure and strong-correlation effects leads to the stabilization of the 
various electronic phases with decreasing temperature.
\end{abstract}

\pacs{71.30.+h, 71.15.Mb, 71.10.Fd, 75.30.Cr}
\maketitle

\section{Introduction\label{intro}}

Since its first characterization~\cite{gardner_actacry_1969} in 1969, the 
understanding of the complex electronic structure of the vanadium sulfide 
\bavs3 poses a longstanding problem in condensed matter physics 
\cite{massenet_jpcsol_1979,mat91,graf_bavs3_pressure_prb_1995,boo99,whangbo_bavs3_puzzling_jsschem_2003}. Numerous experimental and theoretical studies 
have revealed a delicate coupling between orbital, spin and lattice degrees
of freedom over a wide temperature range. On cooling, \bavs3
exhibits three continuous phase transitions, starting with a structural
hexagonal-to-orthorhombic transition at $T_{\rm S}$$\sim$240 K in the metallic
regime. The latter vanishes at $\sim$70 K where a  metal-to-insulator transition 
(MIT), accompanied by a lattice transformation from orthorhombic to 
monoclinic\cite{inami_bavs3_prb_2002,fag05}, to a still paramagnetic phase takes 
place. A final magnetic transition marking the onset of an incommensurable 
antiferromagnetic order~\cite{nak00,hig02} takes place at $T_{\rm X}$$\sim$ 30 K.

The underlying driving forces for these transitions and the specific nature of
the respective phases is to a large extent still a matter of debate. 
It was shown~\cite{forro_bavs3_qcp_prl_2000} that the MIT may be driven to 
zero temperature at high pressure, and the supression of the insulating phase 
leads to non-Fermi-liquid and quantum-critical behavior~\cite{Bar06}. This 
observation is not only adding even more complexity to the already existing 
problems, but also underlines the tricky nature of the electronic structure.
At ambient pressure the MIT is early announced by strong precursive behavior 
such as a large increase of the Hall coefficient~\cite{boo99} and a wide 
one-dimensional (1D) lattice-fluctuation regime~\cite{fagot_bavs3_prl_2003} 
along the $c$ axis of the system. In fact, it seems to be 
established~\cite{inami_bavs3_prb_2002,fagot_bavs3_prl_2003,fag05,lec06} that 
the MIT may be described in terms of a charge-density wave (CDW) instability. 
However, \bavs3 is not a textbook Peierls system. The dc conduction anisotropy 
is rather small~\cite{mih00} ($\sigma_c/\sigma_a$$\sim$3-4), and the ``metallic'' 
phase above \Tmit displays a high resistivity (a few m$\Omega$cm) and 
metallic-like behavior ($d\rho/dT$$>$$0$) only above a weak minimum at 
$\sim$$150$~K, below which it increases upon further 
cooling~\cite{graf_bavs3_pressure_prb_1995,mih00}. Moreover, local-moment 
behavior is revealed from the magnetic susceptibility, with an effective moment 
of approximately one localized spin-$1/2$ per two V sites.  At \Tmit, the 
susceptibility rapidly drops, and the electronic entropy is strongly 
suppressed~\cite{ima96}.

Because of the nominal V$^{4+}$ valence, \bavs3 belongs to the family of 
$3d^1$ compounds with a \t2g manifold spanning the low-energy sector. In the 
hexagonal phase the latter consists (per V ion) of an \A1g and two 
degenerate $E_g$ states. The remaining $e_g$ states are strongly hybridized with 
the S($3p$) states and have major high-energy weight. Below $T_{\rm S}$ the 
degeneracy of the $E_g$s is lifted in the orthorhombic phase. In both phases the 
primitive cell includes two formula units, where the V and S ions form chains 
of face-sharing VS$_3$ octahedra along the $c$ axis. The intrachain V-V distance 
is less than half the interchain distance. It follows~\cite{massenet_jpcsol_1979} 
that the \A1g orbital is mainly directed along the chain, forming a broader band 
due to the significant overlap of neighboring intrachain V ions. On the contrary,
the lobes of the $E_g$ orbitals point inbetween the sulfur ions, i.e., do not 
hybridize strongly with their environment, leading to comparably narrow bands. 
These simple characterizations hold essentially also for the monoclinic 
insulating phase, yet the primitive cell is doubled and the resulting four V 
ions in the basis are now all inequivalent by symmetry~\cite{fag05,fag06}. 
The CDW mechanism has led to a tetramerization, yet no evident 
charge disproportionation among them was detected~\cite{fag06}. The measured 
charge gap~\cite{nak94,graf_bavs3_pressure_prb_1995,mih00,kez06} of about 40 meV 
is twice as large as the apparent spin gap~\cite{nak97}, pointing once more 
towards the relevance of electronic correlations~\cite{mal03}.

In Ref.~[\onlinecite{lec05,lec06}] the orthorhombic phase above the MIT was 
investigated, and it was argued that strong electronic correlations are 
responsible for a substantial charge transfer within the \t2g states, leading 
also to important
Fermi-surface changes in comparison to a weak-correlation treatment. Here we go
further by tracing the low-energy states of \bavs3 all the way from room 
temperature down to 40 K (just above the final magnetic transition).
Although the local environment of the V site does not change dramatically, the 
electronic structure
appears to be rather sensitive to the temperature changes. This originates
from the subtle balance of kinetic energy versus Coulomb interaction in the
electronic system, a characteristic of strongly correlated materials. In fact,
this vanadium sulfide presents an interesting realistic realization of one of
the basic problems in strongly correlated physics: there is nominally one 
electron in the low-energy sector and two distinct orbital states, one forming 
a broader and two forming narrower bands. Hence depending on temperature, nature
shall find the best compromise between kinetic energy gain and potential cost 
due to mutual Coulomb interaction in this multiorbital scenario.

\section{Theoretical framework}

For the investigation of competing band-structure and many-body effects in 
realistic materials, the 
combination~\cite{anisimov_lda+dmft_1997,lichtenstein_lda+dmft_1998} of 
density functional theory (DFT) and dynamical-mean field theory (DMFT) has 
recently proven to be a powerful approach.

For the DFT part we used the local density approximation (LDA) to the 
exchange-correlation energy. The corresponding calculations were performed with
a mixed-basis pseudopotential code~\cite{mbpp_code}. It uses normconserving
pseudopotentials, and plane waves supplemented with some few non-overlapping
localized functions in order to represent the pseudo crystal wavefunction.

Since the low-energy physics of \bavs3 is dominated by the \t2g states, the
so-called LDA+DMFT calculations were performed for the corresponding three-band 
subset. The latter was derived from the full band structure via the 
maximally-localized Wannier function (MLWF) construction~\cite{mar97,sou01}. 
Hence the local 
orbitals which form the impurity in the DMFT context stem from the associated 
Wannier functions (WFs) and the low-energy LDA hamiltonian $H(\bk)$ is expressed 
with respect to these orbitals~\cite{lec06}. To be specific, by making reference 
to the formalism outlined in Ref.~\onlinecite{lec06}, in all LDA+DMFT 
calculations presented here the set of correlated orbtials $\mathcal{C}$ was 
identical to the set $\mathcal{W}$ of the WFs forming the minimal LDA 
hamiltonian. Hence the impurity Green's function is computed in DMFT for finite
inverse temperature $\beta$ via
\begin{equation}\label{gimpwann}
{\bf G}(\iomn)=\sum_{\bk}\left[(i\omega_n+\mu)\mathbf{\openone}
-H^{({\cal C})}(\bk)-\mathbf{\Sigma}^{({\cal C})}(\iomn)\right]^{-1}\,,
\end{equation}
where  $\omega_n$=$(2n+1)\pi/\beta$ are the Matsubara frequencies and 
${\bf\Sigma}$ is the self-energy matrix for the strongly correlated orbitals.
For the local interacting hamiltonian $H_{\rm int}$ the following 
representation restricted to density-density terms only was used:
\begin{eqnarray}
\label{hubmod}
\hat{H}_{\rm int}&=&U\sum_m\hat{n}_{m\uparrow}
\hat{n}_{m\downarrow}+\frac{U'}{2}
\mathop{\sum_{mm'\sigma}}_{m\ne m'}
\hat{n}_{m\sigma}\hat{n}_{m'\bar{\sigma}}\nonumber\\
&&+\frac{U''}{2}\mathop{\sum_{mm'\sigma}}_{m\ne m'}
\hat{n}_{m\sigma}\hat{n}_{m'\sigma}\quad.
\end{eqnarray}
Here $\hat{n}_{m\sigma}$=$\hat{d}_{m\sigma}^{\dagger}
\hat{d}_{m\sigma}^{\hfill}$, where $m$,$\sigma$ denote orbital and spin index.
The following parametrization of $U'$ and $U''$ has been proven to be
reliable~\cite{cas78,fre97} in the case of \t2g-based systems:
$U'$=$U$$-$$2J$ and $U''$=$U$$-$$3J$. We utilized the quantum-Monte Carlo 
(QMC) formalism after Hirsch-Fye~\cite{hirsch_fye} to solve the impurity problem.

For hexagonal and orthorhombic \bavs3 there are only symmetry-equivalent
V ions in the primitive cell and moreover interatomic correlation effects are
not expected to be of crucial importance. Hence a single-site DMFT approach to
describe the strong-correlation effects was employed for those phases. Thereby
the inverse temperature was always $\beta$=30 eV$^{-1}$ and the number of
time slices equaled 128 for the QMC method. However,
since the paramagnetic insulating regime of the low-temperature monoclinic phase 
is associated with a CDW state, this approximation appears inadequate. We thus
used a cluster formalism (for recent reviews see e.g. 
Ref.~[\onlinecite{bir04,mai05,lic02}]), namely the cellular DMFT (CDMFT)  
approach in a realistic context~\cite{pot04,bie05}. More explicitly, the linear 
cluster formed by the four symmetry-inequivalent V ions, each one decorated with a
three-orbital \t2g multiplet, was identified as the impurity for the CDMFT 
scheme. This amounts to a self-energy matrix ${\bf\Sigma}(i\omega_n)$ that is 
not only off-diagonal in the orbital indices, but also in the site indices 
within the cluster. Note however that intercluster components of the self-energy 
are neglected. The latter fact results in the breaking of translational symmetry 
when computing pair correlations for the end sites of our linear cluster. 
However such a cluster approach should still be sufficient to describe the major 
qualitative changes in the correlated electronic structure originating from the 
CDW instability. Since the undertaken cluster investigation is numerically very
expensive within QMC (to our knowledge one of the largest up to now performed in
the framework of realistic cluster-DMFT) we chose $\beta$=25 eV$^{-1}$ and used 
90 time slices. 

\section{Results}
\begin{figure}[t]
\includegraphics*[width=6cm]{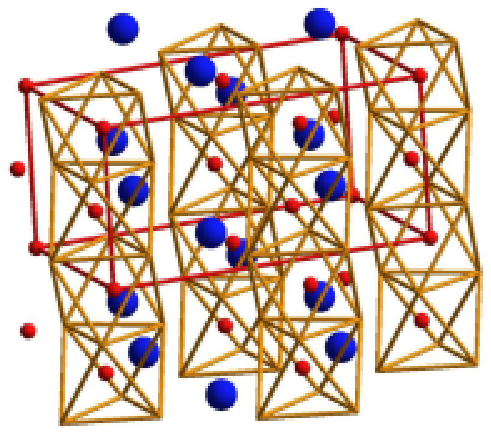}\hspace{0.2cm}
\includegraphics*[width=6cm]{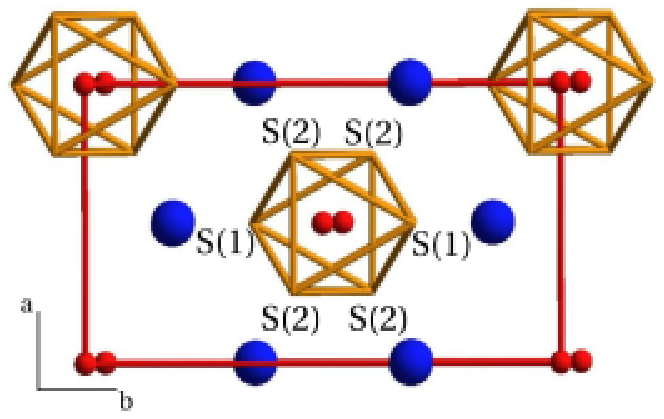}
\caption{(color online) BaVS$_3$ in the $Cmc2_1$ structure. The V ions are 
shown as smaller (red/gray) spheres, the Ba ions as larger (blue/dark) 
spheres.\label{bavs3struc}}
\end{figure}
\begin{table}[t]
\caption{Experimental crystal data used for the investigation of \bavs3.
\label{crysdat1}}
\begin{ruledtabular}
\begin{tabular}{l|l|l|l}
               &  RT           & 100 K        & 40 K \\ \hline\hline
crystal system &  hexagonal    & orthorhombic & monoclinic   \\
space group    &  $P6_3/mmc$   & $Cmc2_1$     & $Im$         \\
$a$ (a.u.)     &  12.71        & 12.77        & 12.78        \\
$b$ (a.u.)     &  22.01$\,(=a\sqrt{3})$   & 21.71 & 21.65        \\
$c$ (a.u.)     &  10.63        & 10.58        & 21.15        \\
$\beta$ ($^0$) &    $-$        &   $-$        & 90.045       \\     
\hline 
experiment     & Ref.~\onlinecite{ghedira_bavs3_neutrons_jpc_1986}  & 
Ref.~\onlinecite{ghedira_bavs3_neutrons_jpc_1986} & 
Ref.~\onlinecite{fag05} \\ 
\end{tabular}
\end{ruledtabular}
\end{table}

Stoichiometric \bavs3 transforms on cooling successively to crystal systems with
lower symmetry, giving a hint to the generally low ordering energy. 
Table~\ref{crysdat1} summarizes the basic crystal data for the three different
phases that are studied in this work. Both, the hexagonal and orthorhombic
phases are associated with the metallic regime, while the monoclinic phase
corresonds to the insulating system. Here we only investigated the 
paramagnetic phase of the insulator and excluded the magnetically 
ordered phase below $T_{\rm X}$.

\subsection{LDA study of the metallic regime}

At room temperature (RT) \bavs3 crystallizes in the hexagonal ($P6_3/mmc$) 
structure~\cite{gardner_actacry_1969} with two formula units in the primitive 
cell. All symmetry operations of the hexagonal group apply to this structure, 
and all Ba, V and S ions in the cell form one single symmetry class, 
respectively. The V ions within the chains are aligned straightly. As noted
by Mattheis~\cite{mattheiss_bavs3_1995}, the variable parameter $x$(S) which 
determines the inplane S-S distances deviates slightly from the ideal hexagonal
value, i.e., $x$(S)=0.1656 while $x_{\rm ideal}$=1/6. This means that the inplane 
equilateral S-S-S triangle above and below an V ion is contracted and the 
intrachain S-S bond lengths are decreased in comparison to the interchain lengths.

\begin{figure}[b]
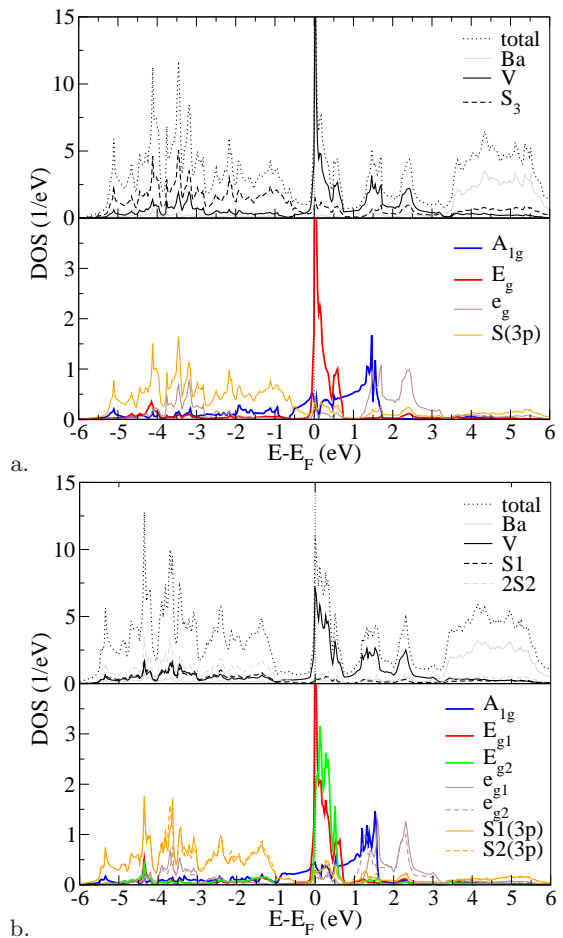

a.\includegraphics*[width=7cm]{DOS.hexa.eps}\hspace{0.5cm}
b.\includegraphics*[width=7cm]{DOS.ortho.eps}
\caption{(color online) Total and local LDA DOS for BaVS$_3$ in (a) the 
$P6_3/mmc$  structure and (b) the $Cmc2_1$ structure.\label{bavs3dos}}
\end{figure}
\begin{figure}[t]
\includegraphics*[width=7cm]{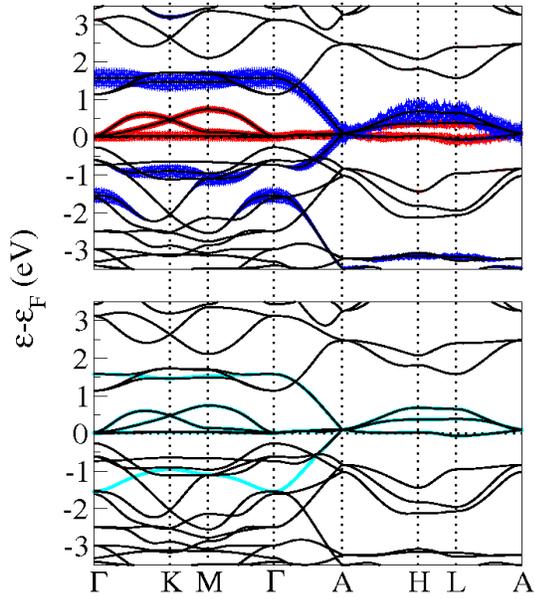}
\caption{(color online) (top) LDA band structure and (bottom) derived 
\t2g Wannier bands for $P6_3/mmc$-BaVS$_3$.  Color coding for (a): 
\A1g (blue), $E_{g1}$/$E_{g2}$ (red).
\label{bavs3hexbands}}
\end{figure}
\begin{figure}[b]
\includegraphics*[width=7cm]{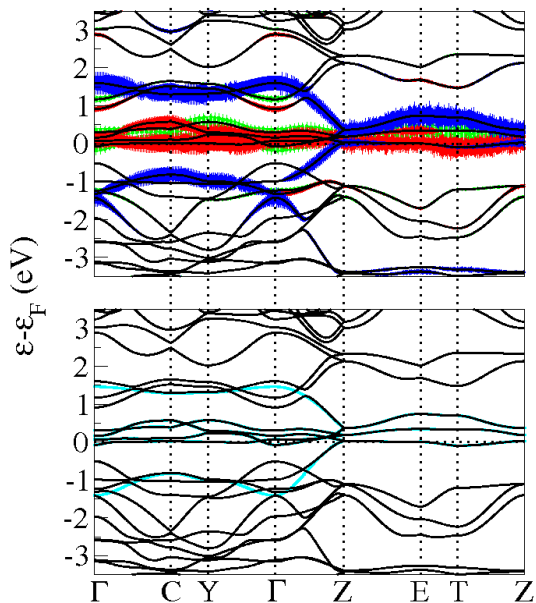}
\caption{(color online) (top) LDA band structure and (bottom) derived 
\t2g Wannier bands for $Cmc2_1$-BaVS$_3$. Color coding for (a): 
\A1g (blue), $E_{g1}$ (red) and $E_{g2}$ (green).
Fig.~\ref{bavs3hexbands}
\label{bavs3orthobands}}
\end{figure}
At $T_{\rm S}$ the crystal system transforms from hexagonal to orthorhombic,
leading to a structure with $(Cmc2_1)$ space group. The V chains are now
zigzag distorted in the $bc$ plane of the lattice. Although still two formula
units form the primitive cell, the symmetry class of the sulfur ions has split
into two: both S1 ions are positioned at (4a) apical sites on the $b$ axis,
while the four S2 ions occupy (8b) sites (see Fig.~\ref{bavs3struc}). From 
Tab.~\ref{crysdat1} it is seen that the structural transformation results in only
minor changes in the lattice parameters, with a maximum 1.5$\%$ per cent 
contraction of the $b$ axis. There are several LDA studies for \bavs3 above the 
MIT~\cite{nak94,mattheiss_bavs3_1995,whangbo_bavs3_jsschem_2002,lec05}. In 
addition to those, a direct comparison of the LDA low-energy electronic structure
for the hexagonal and orthorhombic phases via Wannier construction for the \t2g 
manifold is presented here.

Figure~\ref{bavs3dos} displays the LDA density of states (DOS) for the two
crystal structures. In each case the dominance of the \t2g states at low energy,
with a prominent peak right at the Fermi energy, is evident. For the \A1g 
band the dispersion is indeed reminiscent of 1D characteristics, however below 
the Fermi energy features are changed due to hybridization with the S$(3p)$ 
states. The weight of the latter states is reduced in the energy range $[-1,0]$
for the $Cmc2_1$ structure compared to the $P6_3/mmc$ structure. One might 
interpret this as some decoupling of \A1g and S$(3p)$ throughout the
structural transition. The very large DOS at the Fermi energy, common to both 
structures, due to the $E_g$ states render instabilities towards broken-symmetry
phases very likely.

\begin{figure}[b]
a.\includegraphics*[width=2cm]{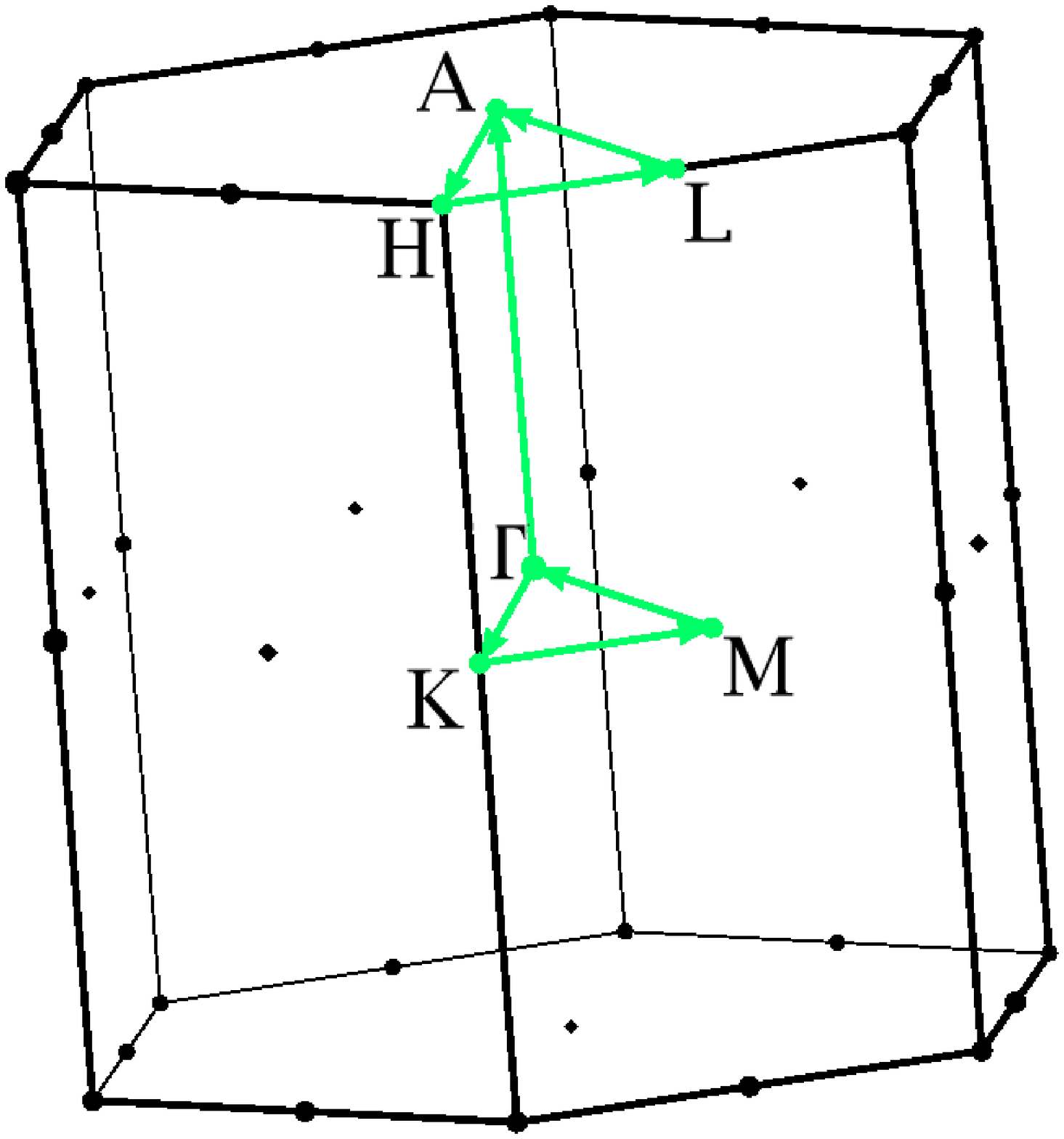}\hspace{0.5cm}
\includegraphics*[width=2cm]{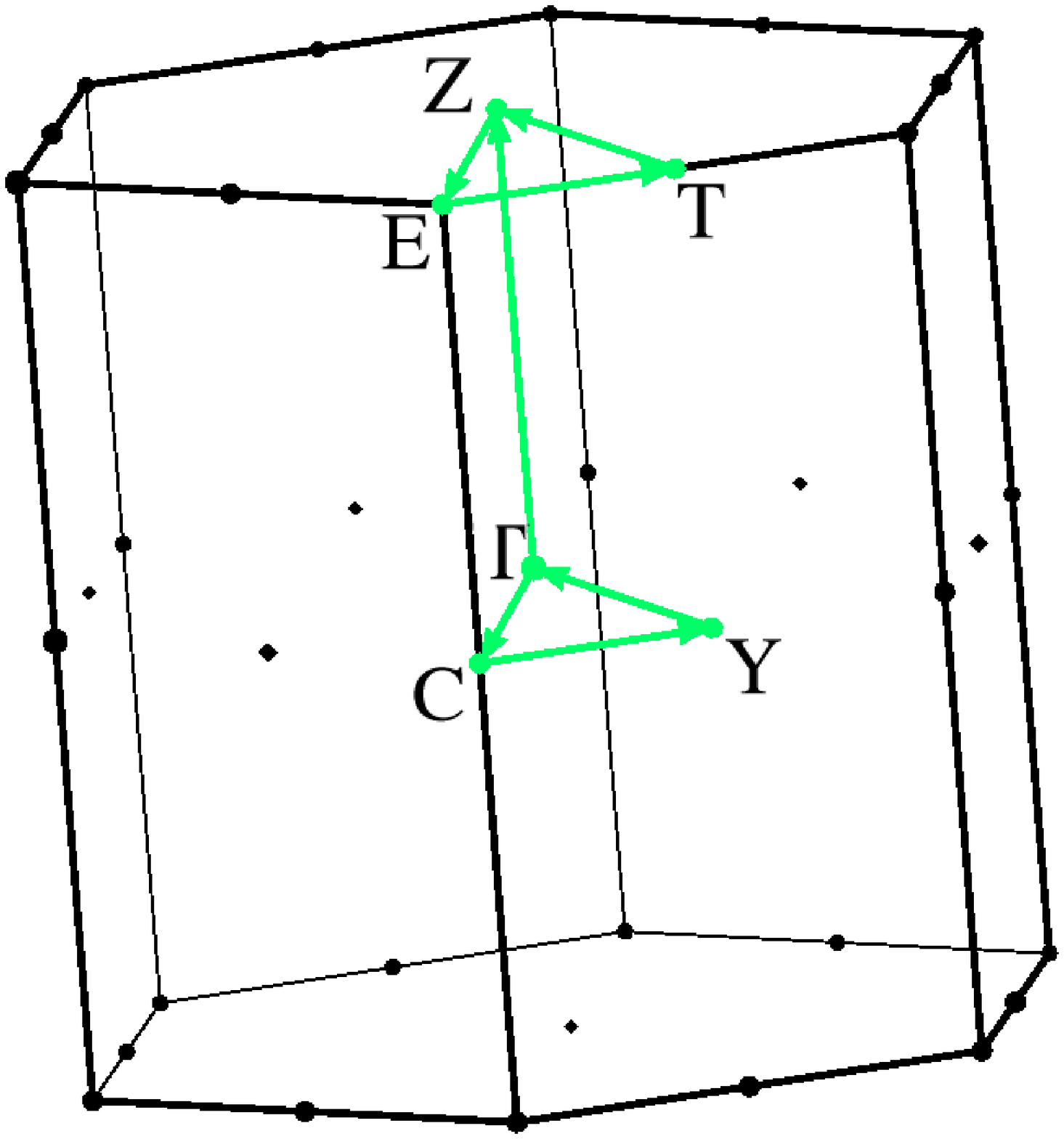}\\[0.4cm]
b.\includegraphics*[width=7cm]{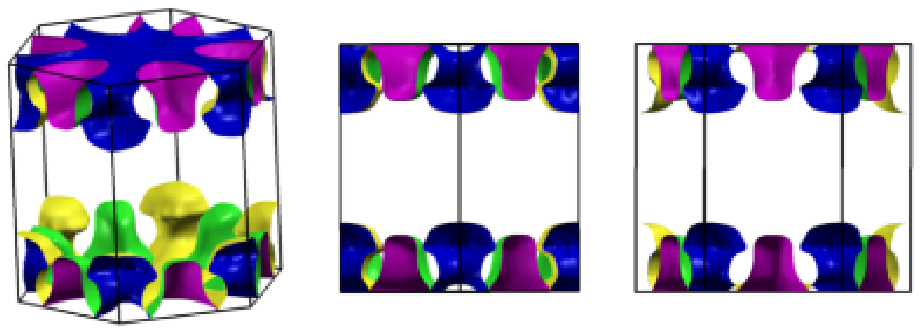}\\[0.2cm]
c.\includegraphics*[width=7cm]{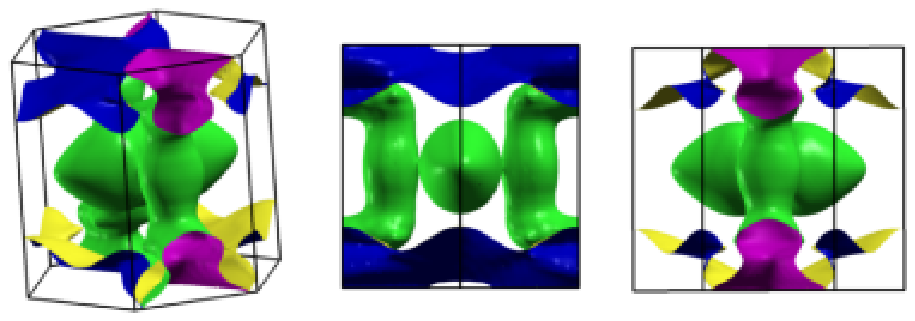}
\caption{(color online) (a) hexagonal and orthorhombic Brillouin zones as
well as LDA Fermi surface for BaVS$_3$ in (b) the $P6_3/mmc$ structure and 
(c) the $Cmc2_1$ structure from different perspectives.\label{bavs3fs}}
\end{figure}
The LDA band structures in Fig.~\ref{bavs3hexbands}a and 
Fig.~\ref{bavs3orthobands}a, with ``fatbands'' exhibiting
the weight of the respective \t2g states on the different bands~\cite{lec06},
show the folded character of the \A1g band due to the two-formula unit
primitive cell. The \A1g bandwidth is dominated by the dispersion along
$\Gamma$-(A,Z), i.e., along the $c^*$ axis. While for the higher-symmetry 
$P6_3/mmc$ structure the folding propagates gapless through A, the corresponding 
bands are separated at Z for the $Cmc2_1$ structure. This hybridization 
between \A1g and $E_g$ is obvious in the upper triangle A-H-L and Z-E-T.
Clearly seen is the hybridization of \A1g with S$(3p)$ resulting in 
``jumps'' of the \A1g character between different bands in the lower
triangle $\Gamma$-K-M and $\Gamma$-C-Y. For both structures the \A1g band 
cuts the Fermi level $\varepsilon_{\rm F}$ close to the zone boundary,  
leading to a nearly filled lower part of the folded band complex. It was noted
in former works~\cite{mat91,whangbo_bavs3_puzzling_jsschem_2003} that for $Cmc2_1$ 
this filling renders a CDW instability within the \A1g band along $\Gamma$-Z 
impossible.

The $E_g$ states form very narrow, in some regions even nearly dispersionless, 
bands right at $\varepsilon_{\rm F}$. An important difference occurs between 
the Fermi surface (FS) of the two metallic phases (see Fig.~\ref{bavs3fs}).
Although in both cases the FS consists of two sheets, the one for $P6_3/mmc$
is entirely located at the zone boundaries with a dominant \A1g sheet and
smaller $E_g$ pockets around L. On the contrary the FS for $Cmc2_1$ shows
as the first sheet a substantial $E_{g2}$ electron pocket centered at $\Gamma$ 
and $E_{g1}$ pillar-like structures on the $b^*$ axis extending along $c^*$. The
FS is completed by the \A1g sheet, now extending deeper into the Brillouin
zone (BZ). Despite the latter observation, this quasi-1D sheet is neither 
strongly flattened nor have both parts the proper distance for nesting
with the experimental CDW vector (${\bf q}^{\rm (exp)}_c$=$0.5{\bf c}^*$).
\begin{figure}[t]
a.\includegraphics*[width=7cm]{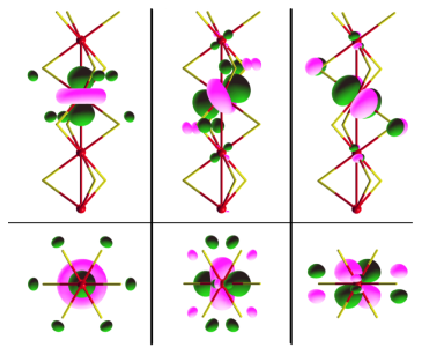}\\[0.35cm]
b.\includegraphics*[width=7cm]{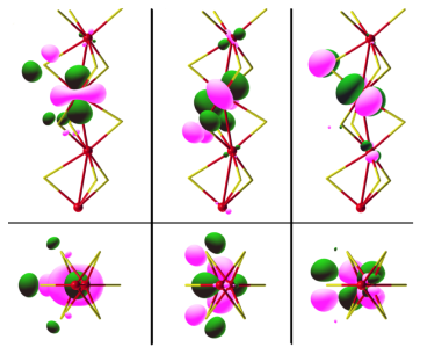}
\caption{(color online) \t2g Wannier functions for BaVS$_3$ in the 
crystal-field basis of (a) the $P6_3/mmc$ structure and (b) the $Cmc2_1$ 
structure (b).\label{bavs3wannier}}
\end{figure}

The dispersions according to the derived three-band hamiltonian on the basis
of the maximally-localized procedure~\cite{mar97,sou01} are shown in 
Fig.~\ref{bavs3hexbands},\ref{bavs3orthobands}b. Because of the entanglement 
of the V($3d$) bands with the S($3p$) ones, the dominant \A1g-like band does 
not coincide with the true LDA bands in this minimal model~\cite{lec06}. Still 
such a three-band approach should carry the essential physics in the low-energy
regime. 
Figure~\ref{bavs3wannier} pictures the corresponding V(\t2g) Wannier orbitals in 
the crystal-field basis, i.e., with vanishing onsite hybridization. For both
phases it is seen that the $E_{g1}$ orbitals leak out on the S2 ions, while 
the $E_{g2}$ orbitals have weight on the apical S1 atoms (recall that (S1,S2)
are symmetrically equivalent only in the hexagonal phase). The \A1g orbital
hybridizes with the $3p$ orbitals on both sulfur-ion types. That this 
hybridization is indeed weakened in the $Cmc2_1$ structure may be derived from 
the reduced spread of the \A1g WF shown in Tab.~\ref{table-spread}. 
Interestingly, the spread is now even below the values for the $E_g$ states. Note
also that the WF centers are identical with the V positions only for the 
hexagonal structure, while there are some shifts for the orthorhombic structure. 
The main qualitative difference in the hopping integrals (Tab.~\ref{table-bahop})
between the two structures is the emerging substantial hybridization between 
\A1g and $E_{g1}$ for $Cmc2_1$. Note that generally the $E_g$ hoppings are rather
isotropic, whereas the dominating hopping along the $c$ axis in the case of \A1g
is obvious.
\begin{table}[t]
\caption{Wannier centers $\bR_w$ and spread $\langle r^2\rangle$ of the 
\t2g-like MLWFs constructed from a (6$\times$6$\times$6) $k$-point mesh. The
positions of the symmetrically equivalent V sites in cartesian coordinates 
(in a.u.) read for $P6_3/mmc$: $\bR_{\rm{V1}}$=(0.00,0.00,0.00) and
$\bR_{\rm{V2}}$= (0.00,0.00,5.30) ; as well as for $Cmc2_1$: 
$\bR_{\rm{V1}}$= (0.00,0.46,-0.01) and $\bR_{\rm{V2}}$= (0.00,-0.46,5.28). 
The V(2) site is symmetry-related to the V(1) site by the symmetry operation 
$C_2^{(z)}\bR_{\rm{V1}}$+0.5. In the following only the data for V1 is 
shown.\label{table-spread}}
\begin{ruledtabular}
\begin{tabular}{l|c|c|c}
structure & WF & $\bR_w-\bR_{\rm{V1}}$ (a.u.) & $\langle r^2\rangle$ (a.u.$^2$) \\ \hline
           & \A1g     & 0.00,  0.00, 0.00                & 18.62    \\
$P6_3/mmc$ & $E_{g1}$ & 0.00,  0.00, 0.00                & 17.10  \\
           & $E_{g2}$ & 0.00,  0.00, 0.00                & 17.10 \\ \hline
           & \A1g     & 0.00,  0.30, -0.16               & 16.60   \\
$Cmc2_1$   & $E_{g1}$ & 0.00,  0.19,\hspace{0.1cm}  0.35 & 17.55 \\
           & $E_{g2}$ & 0.00,  0.56, -0.31               & 17.53 \\
\end{tabular}
\end{ruledtabular}
\end{table}
\begingroup
\squeezetable
\begin{table}[t]
\caption{Hopping integrals between the \t2g Wannier orbitals of \bavs3 in the
crystal-field basis. The first value corresponds to the $P6_3/mmc$ structure 
and the second to the $Cmc2_1$ structure, respectively. The term 
'00$\frac{1}{2}$' shall denote the hopping to the nearest-neighbor V site 
within the unit cell. One of the nearest-neighbor V ions in the $ab$ plane is 
located at '100', while '110' and '$\bar{1}$10' are closest V ions along $a$ and
$b$, respectively. Energies in meV.\label{table-bahop}}
\begin{ruledtabular}
\begin{tabular}{c|rr|rr|rr|rr|rr|rr}
           & \A1g & \A1g  & $E_{g1}$ & $E_{g1}$ & $E_{g2}$ & $E_{g2}$ & \A1g & $E_{g1}$ & \A1g & $E_{g2}$ & $E_{g1}$ & $E_{g2}$ \\ \hline
000            & 395 & 423 & 200 & 210 & 200 &236 & 0 &    0 & 0 & 0 &  0 & 0 \\
00$\frac{1}{2}$&-587 &-511 &  90 &  44 & -90 &-12 & 0 & -146 & 0 & 0 &  0 & 0 \\
001            & -61 & -86 &   5 &  14 &   5 &  3 & 0 &    7 & 0 & 0 &  0 & 0 \\
100            & -49 & -35 &   8 &  14 & -40 &-26 & 18 &-7 & -32 &-14 & 41& 14 \\
110            & -49 & -26 & -63 & -76 &  31 & 29 & -37 &-28& 0 & -2 & 0 &-12 \\
$\bar{1}$10    &   3 &   1 &   0 &   2 & -5 & -6 & 0 & -2 & 0 &  0 & 0&  0 \\
\end{tabular}
\end{ruledtabular}
\end{table}
\endgroup

\subsection{LDA+DMFT study of the metallic regime}

The LDA approach presented in the last section underlies the assumption that the
mutual interactions between the electrons may be cast into a static local
exchange-correlation potential within an effective single-particle description.
However it is known that for strongly correlated systems such a description is 
likely to fail. Since one indeed expects rather strong correlations within the 
quarter-filled V(\t2g) states of \bavs3, also due to the specific 
characteristics of very narrow $E_g$ bands and the broader \A1g band, we employed 
the LDA+DMFT framework to explicitly include many-body effects in the electronic
structure.

\begin{figure}[b]
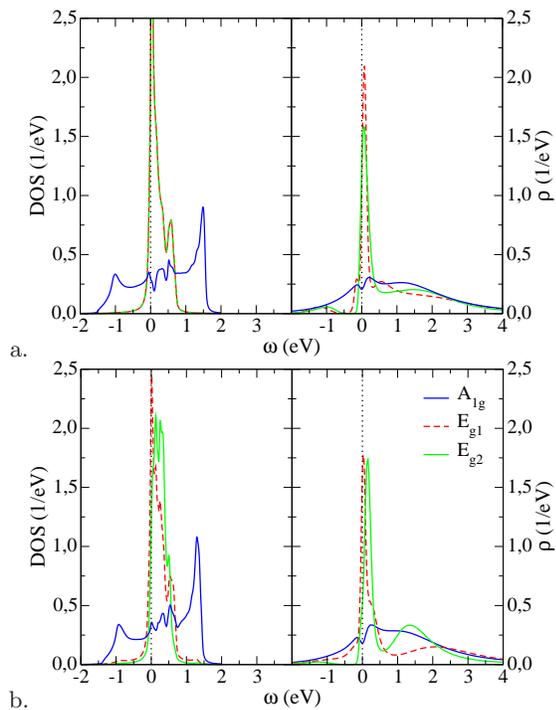

a.\includegraphics*[width=7cm]{bavs3-specfunc.hexa.eps}\hspace{0.5cm}
b.\includegraphics*[width=7cm]{bavs3-specfunc.ortho.eps}
\caption{(color online) 
(right) Local spectral functions from LDA+DMFT in comparison to (left) the 
local LDA DOS for the \t2g WFs in the crystal-field basis, for (a) the 
$P6_3/mmc$ structure and (b) the $Cmc2_1$ structure. The QMC solver was used 
for $T$=390 K.
\label{bavs3spec}}
\end{figure}

By identifying the derived \t2g WFs in the crystal-field basis as the subspace 
of correlated orbitals, we derived the $\bk$-integrated (local) spectral 
functions $\rho(\omega)$ shown in Fig.~\ref{bavs3spec}. Clearly seen is the 
transfer of spectral weight from the quasiparticle (QP) peaks to lower/higher 
energies in comparision to the local LDA DOS, especially for the $E_g$ states. 
This corresponds to an inclusion of the atomic-like excitations important for 
states with substantial localized character, which is missing in the standard 
LDA picture. By varying $T$ one observes~\cite{lec05} additionally that the 
strength of the $E_g$ QP peak changes also more significantly. Thus the 
corresponding electrons are effectively localized for a wide temperature range 
due to incoherence effects.

Since the correlation effects influence the subtle energetic balance in this 
system, they are moreover responsible for a substantial charge transfer between 
the relevant orbitals~\cite{lec05,lec06}, resulting in different orbital-resolved 
fillings compared to the LDA ones (see Tab.~\ref{table-fillings}).
The LDA filling of the \A1g band is close to 70\% in the hexagonal structure
and does not change much for temperatures where the DMFT study was elaborated.
A reasonable choice~\cite{lec05} for the Hubbard 
parameters $U$ and $J$ leads to a significant transfer of charge from \A1g to
$E_g$, in order the overcome the large potential energy cost for occupying mainly
the former orbital. The new fillings for $E_{g1}$ and  $E_{g2}$ differ slightly
due to a marginal hybridization between \A1g and $E_{g1}$ in the Wannier 
hamiltonian already for the hexagonal structure. Note that {\sl filling} in
LDA+DMFT is not equivalent to pure band filling in the LDA sense, because the 
atomic-like excitations are now also included. The essential change for 
the orthorhombic structure below $T_{\rm S}$ (which was here treated at same $T$
within the QMC solver of DMFT) is the effective reduction of the three-band
to a dominant two-band problem. Due to the now substantial \A1g-$E_{g1}$
hybridization the charge transfer is dominantly taking place between those two
orbitals. Yet the overall occupation of the \A1g orbital is only little smaller
than in the hexagonal phase. The $E_{g1}$ filling is now close to 50\%, in 
good agreement with the experimentally observed local magnetic moment of about
one free spin every other V ion.

\begin{table}[t]
\caption{Orbital-resolved fillings for \bavs3 from LDA+DMFT within the 
crystal-field Wannier basis. The QMC solver was used for $T$=390 K.
\label{table-fillings}}
\begin{ruledtabular}
\begin{tabular}{l|r|ccc}
structure & $U$,$J$ (eV) & $A_{1g}$ & $E_{g1}$ & $E_{g2}$ \\ \hline
                & 0.0, 0.0 & 0.67 & 0.16 & 0.16 \\
\bw{$P6_3/mmc$} & 3.5, 0.7 & 0.45 & 0.29 & 0.27 \\ \hline
                & 0.0, 0.0 & 0.59 & 0.31 & 0.10 \\
\bw{$Cmc2_1$}   & 3.5, 0.7 & 0.41 & 0.45 & 0.14 \\
\end{tabular}
\end{ruledtabular}
\end{table}
\begin{figure}[b]
a.\includegraphics*[width=7cm]{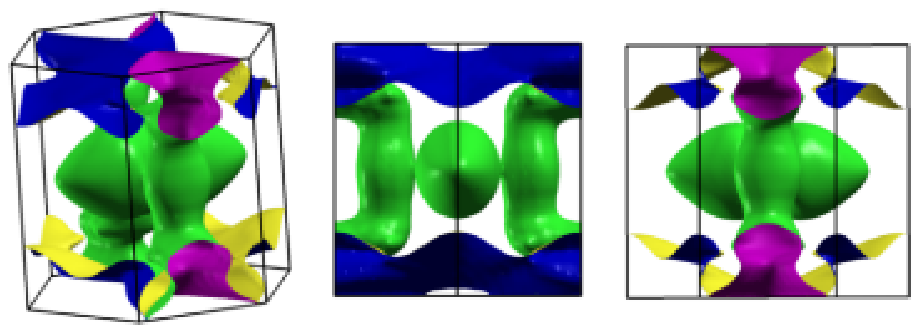}\\[0.2cm]
b.\includegraphics*[width=7cm]{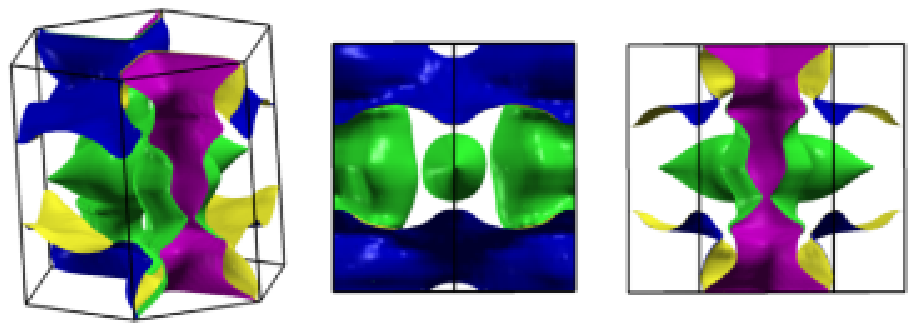}
\caption{(color online) comparison between (a) the LDA FS of orthorhombic \bavs3
and (b) the corresponding QP FS derived from LDA+DMFT.\label{bavs3fsdmft}}
\end{figure}

Besides the orbital-resolved filling, the change of the respective FS 
sheets in the metallic regime is also of large interest. However keep in mind 
that there is no straightforward relation between those two issues, as only the
{\sl total} Fermi-surface volume is invariant when turning on correlations and 
no unique rule of how the individual sheets have to change can be derived. But
this change can of course be calculated, which was done for the orthorhombic
phase with the Wannier basis in Ref.~\onlinecite{lec06}, and the change of the 
FS sheets is indeed in line with what one expects from the overall charge 
transfers (s. Fig.~\ref{bavs3fsdmft}). More explicitly, a strong Fermi-surface 
deformation was revealed, placing parts of the renormalized \A1g sheet now in 
reasonably good position for a possible nesting with the experimentally 
determined ${\bf q}$ vector. An important observation was that the nesting 
should mainly take place away 
from the high-symmetry directions in the BZ, i.e., the \A1g band along the 
$\Gamma$-Z direction should not be strongly involved in the direct nesting. As 
pointed out, one expects the $E_g$ states to be essentially localized for higher 
temperatures, rendering the definition of a sharp FS rather difficult. For this 
reason, no renormalized FS was computed for the hexagonal phase. Due to the 
missing \A1g-$E_{g1}$ hybridization one would however expect that for this phase 
the overall \A1g sheet is shifted more or less coherently in the BZ, contrary to 
the orthorhombic case.

\subsection{LDA study of the paramagnetic insulating regime}

\begin{figure}[b]
\includegraphics*[height=5cm]{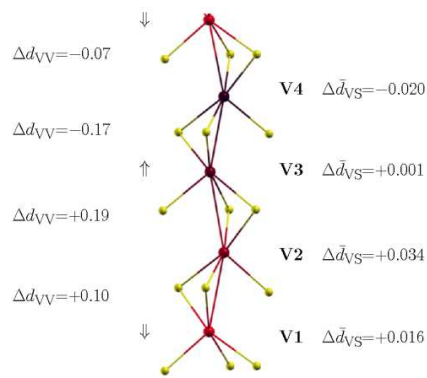}
\caption{(color online) Single \bavs3 chain for the monoclinic structure at 40 K.
The numbers on the left mark the deviation of intrachain V-V distance in 
comparison to the value $d_{\rm VV}$=$5.37$ a.u. in $Cmc2_1$-BaVS$_3$, whereas 
the numbers on the left denote the deviation of the average V-S distance. All 
values in a.u.. The arrows indicate the dominant shifts of the ions again with 
respect to the orthorhombic structure above the CDW instability. 
\label{bavs3monostruk}}
\end{figure}

\bavs3 below \Tmit is insulating with a monoclinic $Im$ structure involving four 
formula units in the primitive cell~\cite{fag05}. The system is described as
a CDW state with a dominant $2k_F$ distortion~\cite{fag05}.
Figure~\ref{bavs3monostruk} 
displays the four inequivalent vanadium ions along the chain together with 
numbers indicating the shift of the atomic positions with reference to the 
$Cmc2_1$ structure above the MIT (at $T$=100 K). It is seen that the shifts for
V along the chain are relatively small, on the scale of 4\% at most. Nonetheless,
a dominant disortion pattern for this CDW state may be identified. The mainly
shifted ions are V1 and V3 in our notation, whereas V2 and V4 only marginally
change their positions. Since V1 and V3 are shifted towards each other, with
V4 inbetween, the tetramerization appears as an effective trimerization, 
isolating the V2 site. As a result, the V-V distances in decreasing order are:
V2-V3, V1-V2, V1-V4 and V3-V4. To a smaller extent, the average V-S distance
(taking into account only the six nearest-neighbor sulfur ions, respectively)
also varies. Again, the averaged distance $\Delta\bar{d}_{\rm VS}$ is largest
for the V2 ion and smallest for the V4 ion.

\begin{figure}[t]
\includegraphics*[width=7cm]{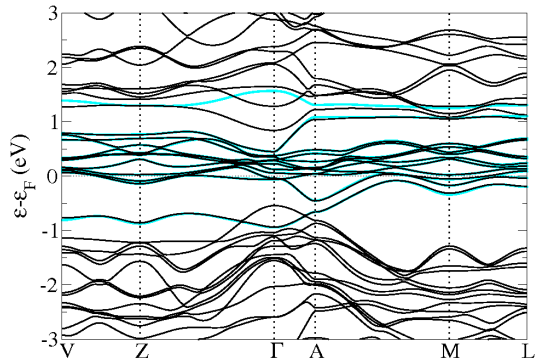}
\caption{(color online) LDA band structure and derived \t2g Wannier 
bands (cyan/lightgray) for monoclinic $Im$-BaVS$_3$.  
\label{bavs3monobands}}
\end{figure}
Although the CDW phase is experimentally 
known~\cite{graf_bavs3_pressure_prb_1995} to be paramagnetic insulating, the
LDA calculation finds it to be metallic (see Fig.~\ref{bavs3monobands}). Hence
whereas the insufficiency of LDA for the metallic regime required a closer band-structure study, the failure due to the neglection of strong correlations 
is now obvious. Still the LDA approach may deliver relevant information about 
the kinetic part of the hamiltonian and the changes of the electronic structure 
due to the CDW distortion. 

The total LDA DOS for the $Im$ structure is rather close to the $Cmc2_1$ one. We 
plotted in Fig.~\ref{bavs3monobands} only the Wannier bands for the low-energy 
regime on top of the 
full LDA band structure. The lowest and highest \t2g bands still have the 
strongest \A1g weight, and the hybridization between \A1g and $E_{g1}$ is 
significant especially at higher energy. However no clear distinction between 
the individual electronic character of the V ions can be made on the level of 
simple projection onto local orbitals. 

\begin{figure}[t]
\includegraphics*[width=7cm]{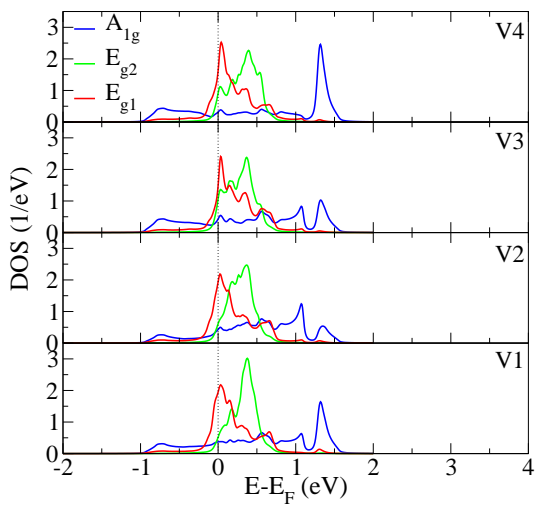}
\caption{(color online) LDA DOS based on the derived \t2g Wannier 
functions for monoclinic $Im$-BaVS$_3$.  
\label{bavs3monodos}}
\end{figure}
\begin{table}[t]
\caption{Onsite terms of the \t2g Wannier hamiltonian for $Im$-\bavs3 compared
to the ones in the metallic phases. The values are in meV.\label{table-cfsplit}}
\begin{ruledtabular}
\begin{tabular}{c|c c c c|c|c}
          & V1  & V2  & V3  & V4 & $Cmc2_1$ & $P6_3/mmc$\\ \hline
 \A1g     & 478 & 473 & 460 & 477 & 423 & 395 \\
 $E_{g1}$ & 431 & 431 & 436 & 432 & 210 & 200 \\
 $E_{g2}$ & 448 & 444 & 443 & 448 & 236 & 200 \\ 
\end{tabular}
\end{ruledtabular}
\end{table}
\begin{table}[t]
\caption{Spread $\langle r^2\rangle$ of the \t2g WFs for $Im$-\bavs3 compared to 
the ones in the metallic phases. The values are in (a.u.$^2$).
\label{table-imspread}}
\begin{ruledtabular}
\begin{tabular}{c|c c c c|c|c}
          & V1  & V2  & V3  & V4 & $Cmc2_1$ & $P6_3/mmc$\\ \hline
 \A1g     & 22.72 & 21.13 & 22.18 & 23.69 & 16.60 & 18.62 \\
 $E_{g1}$ & 17.61 & 18.04 & 18.39 & 17.54 & 17.55 & 17.10 \\
 $E_{g2}$ & 18.86 & 19.31 & 17.91 & 18.21 & 17.53 & 17.10 \\ 
\end{tabular}
\end{ruledtabular}
\end{table}

A bit more insight is obtained when going to the Wannier represenation, again 
using the crystal-field basis as the choice of reference. 
Figure~\ref{bavs3monodos} indicates that from a low-energy perspective the
(V1,V2) and (V3,V4) ions have similar characteristics, especially in the 
occupied part of the DOS. The \A1g occupation compared to the $E_{g1}$ one is 
larger for the (V3,V4) ions. Because of the way of the shifts of the atomic 
positions in the monoclinic structure it is not so surprising to find that 
already on an LDA level the (V3,V4) and (V1,V2) ions appear to form somehow 
two different classes. Since the (V1,V2) ions are more isolated, the larger 
$E_{g1}$ filling makes sense when being the more localized state. 
In spite of these differences it is seen in Tab.~\ref{table-cfsplit} that the 
LDA crystal-field splitting within the \t2g manifold is greatly reduced in the CDW
state. Note also that the spread of the \A1g WF is significantly enhanced in the
LDA description of $Im$-\bavs3 (see Fig.~\ref{table-imspread}).

\subsection{LDA+DMFT study of the paramagnetic insulating regime}

\begin{figure}[b]
\includegraphics*[width=7cm]{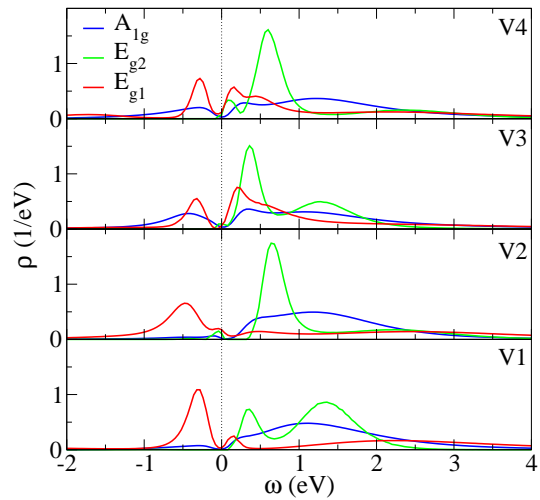}
\caption{(color online) Local spectral functions from LDA+CDMFT
for monoclinic $Im$-BaVS$_3$. The QMC solver was used at $T$=460 K.  
\label{bavs3monospec}}
\end{figure}
In order to overcome the obvious failure of LDA in describing the paramagnetic
insulating state of \bavs3 an LDA+CDMFT approach was employed. Because of
the revealed different behavior of the inequivalent V ions already on the LDA
level, the minimal cluster has to include all four V ions along the chain. Such
a four-site cluster leads in the present case to the description within an 
effective twelve-band model on the basis of the derived Wannier hamiltonian. 
The same values for $U$ and $J$ as for the metallic regime were used ($U$=3.5 eV,
$J$=0.7 eV), no explicit
interatomic Coulomb repulsion term was introduced. Due to the large computational
effort for the QMC impurity solver, the lowest temperature with a still 
reasonable statistics we achieved was $T$=460 K.

\begin{table}[b]
\caption{Orbital-resolved fillings for $Im$-\bavs3 from LDA+CDMFT within 
the crystal-field Wannier basis. The values in the left column 
correspond to the LDA result, while in the right column the values for 
$U$=3.5 eV, $J$=0.7 eV (with $T$=460 K in the QMC impurity solver) are given.
\label{table-imfillings}}
\begin{ruledtabular}
\begin{tabular}{c|l r|l r|l r|l r|l r}
         &  V1  & & V2   & & V3   & & V4   & & $\bar{\mbox{V}}$& \\ \hline

\A1g     & 0.49 & 0.12 & 0.40 & 0.11 & 0.62 & 0.47 & 0.61 & 0.34 & 0.53 & 0.26 \\
$E_{g1}$ & 0.46 & 0.89 & 0.44 & 0.85 & 0.28 & 0.46 & 0.37 & 0.62 & 0.39 & 0.70 \\
$E_{g2}$ & 0.05 & 0.03 & 0.07 & 0.07 & 0.11 & 0.03 & 0.10 & 0.02 & 0.08 & 0.03 \\ \hline
 sum     & 1.00 & 1.04 & 0.91 & 1.03 & 1.01 & 0.96 & 1.08 & 0.98 &      & 
\end{tabular}
\end{ruledtabular}
\end{table}
Figure~\ref{bavs3monospec} exhibits the site- and orbital-resolved local
spectral function obtained from CDMFT. It is seen that within this
description the system can be interpreted to be in an insulating state. 
The partially still remaining minor spectral weight at zero energy is due
to the limitation concerning the handable temperature with the QMC solver. Remember 
that the experimental charge gap from precise optics measurements~\cite{kez06} equals 
only $\Delta_{\rm ch}$=0.42 meV.

Obviously the strong correlations lead to a substantial renormalization of the
crystal-field splitting, since the $E_g$ twofold is now widely separated in 
energy, shifting the $E_{g2}$ state towards higher energy and somehow ``out of
the picture''. The different tendencies in the character of the inequivalent V
ions seen in LDA, are now much more strongly enhanced. Thus the (V1,V2) ions 
have now nearly exclusively $E_{g1}$ weight, while the (V3,V4) ions show some
mixed \A1g/$E_{g1}$ occupation. This interesting result is also summarized in
the site- and orbital-resolved occupations shown in Tab.~\ref{table-imfillings}.
Hence the charge transfer that was observed in the metallic regime due to strong
correlations, takes place also in the insulating state, however now in a 
site-dependent manner. Whereas the (V1,V2) ions loose their \A1g occupation
nearly completely and become orbitally polarized, the (V3,V4) ions fall more in 
the regime of orbital compensation, with the V4 ion gaining some extra $E_{g1}$
weight. On average, roughly speaking the $E_{g1}$ orbital is the winner of the
CDW transition, since it replaces the \A1g orbital as the dominant orbital in the
system. Concerning the question of charge order, the data does not provide a
strong argument for either side. Albeit from the numbers there appears to be the 
slight tendency to put some minor extra charge on the (V1,V2) ions, due to the
restrictions in quantitative accuracy of the formalism this may be within the 
error bars. Remember that experimentally no charge order was found~\cite{fag06}.

The LDA+CDMFT method gives access to more quantities than solely the
spectral functions and onsite densities. To find out more about the nature of
the CDW transition and the insulating state in \bavs3, investigating the behavior
of the self-energy ${\bf \Sigma}(i\omega_n)$ is very instructive. Note that in 
the present case ${\bf \Sigma}$ corresponds to a 12$\times$12 matrix, including 
information not only about onsite but also intersite correlations within the 
four-site cluster.
\begin{figure}[ts]
\includegraphics*[width=7cm]{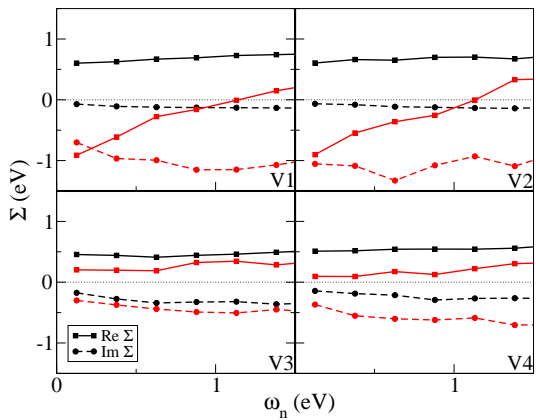}
\caption{(color online) \A1g and $E_{g1}$ onsite self-energy 
$\Sigma(i\omega_n)$ for the four inequivalent V ions in monoclinic 
$Im$-BaVS$_3$ at $T$=460 K. Dark lines correspond to \A1g, red/gray lines to 
$E_{g1}$. \label{bavs3sigmaon}}
\end{figure}
\begin{figure}[b]
\includegraphics*[width=7cm]{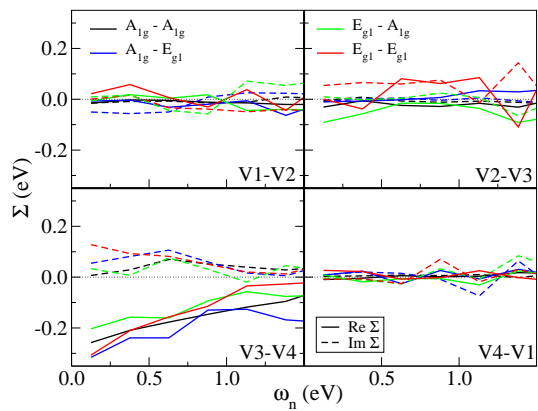}
\caption{(color online) Intersite self-energy $\Sigma_{\rm V-V}(i\omega_n)$
for the nearest-neighbor V-V pairs in monoclinic $Im$-BaVS$_3$ at $T$=460 K. 
Note that the shown $\Sigma_{\rm V4-V1}$ is not a true pair self-energy here
(see text).
\label{bavs3sigmain}}
\end{figure}

It is seen in Fig.~\ref{bavs3sigmaon} that the onsite self-energies for the 
various V ions display the expected behavior. While $\Sigma_{A_{1g}}$ and
$\Sigma_{E_{g1}}$ have rather different amplitude and zero-frequency slope
for (V1,V2), they scale similar for (V3,V4). The large negative increase of 
Re $\Sigma_{E_{g1}}$ close to zero frequency for (V1,V2) leads to the strong 
shift of the QP spectral weight to lower energies observed in 
Fig.~\ref{bavs3monospec}. An important qualitative finding is that none of the 
Im $\Sigma$ diverges at $\omega_n$=0. Hence the opening of the gap in \bavs3
is due to shifts of the QP states away from zero energy.

The inspection of the nearest-neighbor self-energy $\Sigma_{\rm V-V}$ (see
Fig.~\ref{bavs3sigmain}) reveals more details of the CDW state. For the (V3,V4)
ions Re $\Sigma_{\rm V-V}$ displays a salient increasing behavior when 
approaching zero frequency. Such a tendency for the intersite self-energy is a 
strong indication for the importance of intersite correlation effects leading
to interatomic dimer formation~\cite{bie05}. However within the crystal-field 
basis (derived from the LDA hamiltonian) the occupation on (V3,V4) is of mixed 
(\A1g,$E_{g1}$) character. Diagonalizing the interacting cluster Green's 
function thus may lead to an orbital basis which corresponds to this dimer 
symmetry. On the other hand, the V1-V2 pair does not show strong intersite
correlations. Thus these dominant $E_{g1}$ occupied ions do not tend to form a 
spin singlet. It follows therefrom that the apparent spin gap in the insulating
system is not originated from the direct correlated coupling of the neighboring
$E_{g1}$ spins on (V1-V2). However, one observes for the V2-V3 self-energies a 
minor tendency for singular behavior, especially for the $E_{g1}$-\A1g channel.
One may conclude from this that the $E_{g1}$ spins on (V1,V2) are somewhat
effectively pinned by the neighboring dimers within the overall tetramerized
state. This would explain the large drop in the magnetic 
susceptibility~\cite{graf_bavs3_pressure_prb_1995} and 
excess entropy~\cite{ima96} below the MIT. 

It has to be noted that the
intersite $\Sigma_{\rm V-V}$ for V4-V1 is not a true pair self-energy as the
others. Since the cellular DMFT approach we applied here breaks translational
invariance, this special intersite self-energy connecting the surface of the
cluster may not be obtained accurately. Thus for completeness we plotted in 
Fig.~\ref{bavs3sigmain} instead $\Sigma_{\rm V1-V4}$. The issue of 
translational-invariance breaking is surely a drawback of the used method. However
since the V4-V1 pair is expected to be in an intermediate state between V1-V2
and V3-V4, we believe that the qualitative result of how the V ions electronically
relate to each other remains unchanged. Note that one option to restore the
periodicity in the present context would have been to perform calculations in the
chain-DMFT~\cite{bie01} framework. However, there the different treatment of 
inter- and intrachain hopping might cause other problems for \bavs3, since the
1D character (as stated) is not very strongly indicated from the band 
hamiltonian.

\section{Summary and conclusions}
The puzzling physics of \bavs3 is dominated by the competition between the more 
itinerant \A1g state and the quasi-localized $E_{g}$ states, which form together 
the \t2g manifold of the V$(3d)$ shell. Some theoretical models in the 
early days tried to rule out one or the other of those orbital sectors for
playing an essentials role. However several recent experimental and theoretical 
studies revealed clearly the importance of the existence of both orbital types
in order to find a way of understanding the complex electronic phases of this
system. Thus \bavs3 appears to be a manifest multiorbital system, and if at all
some orbital degrees of freedom might just freeze out at very low temperatures.

LDA approaches to the electronic structure of \bavs3 tend quite naturally to
overestimate the itinerant character of the system. The \A1g associated Wannier
orbital has an LDA filling of around 70\% for the RT hexagonal structure, while
the inclusion of strong electronic correlations within the LDA+DMFT reduce this
filling to about 50\%. This is because a dominant \A1g filling becomes just too 
costly in the presence of reasonable mutual Coulomb interactions between the
electrons. The balanced occupation of \A1g and $E_g$ seems to be stable when 
going to the orthorhombic structure below $T_{\rm S}$, yet the LDA+DMFT 
calculations revealed some tendency towards increasing the $E_g$ filling even 
more. Note however that there is still some arbitrariness in what one calls an
\A1g/$E_g$ {\sl orbital}, especially in the metallic regime and between 
different phases, and hence some margin in the derived numbers. Nonetheless,
it was shown that in the paramagnetic insulating phase with the monoclinic
structure the average $E_g$ occupation finally reaches about 70\%. 

The $E_g$ states show a low QP coherence temperature and are
expected to be effectively localized for elevated temperatures in the metal.
Though the LDA approach (even when extended by linear expansions of the DMFT 
self-energy~\cite{lec06}) shows $E_g$ FS sheets, it is very likely that the 
$E_g$ QPs do not participate in the true Fermi surface in a well-defined manner. 
Hence due to the rather large difference in the coherence temperatures, one may 
describe \bavs3 to be in an effective orbital-selective insulating regime. As
there is substantial hybridization between \A1g and $E_{g1}$ below $T_{\rm S}$,
the low metallicity of \bavs3, and especially the bad-metal regime below
150 K, may result from the scattering processes for the 
(quasi-)itinerant electrons. The origin of the hexagonal-to-orthorhombic
transition appears to be closely related to the \A1g-$E_{g1}$ hybridization. It
was shown that the $E_{g1}$ Wannier orbital connects to two S2 ions, while
the $E_{g2}$ one hybridizes with only one apical S1 ion. It is likely that this
imbalance favors the susceptibility for growing symmetry-breaking \A1g-$E_{g1}$ 
hopping. Thereby the large DOS close to the Fermi level is partly reduced when 
lifiting the $E_g$ degeneracy, and this driving force leads via the final zigzag 
intrachain distortion to a new energetic minimum. Albeit many details of the MIT
are still open, as discussed in Ref.~\onlinecite{lec06} the correlation-induced
flattened \A1g FS sheets away from the high-symmetry directions in the BZ are 
good candidates for a meaningful matching with experimental 
findings~\cite{fagot_bavs3_prl_2003}. Further experimental studies of the 
low-energy regime close to the MIT are needed to reveal more details.

We revealed with LDA+CDMFT calculations that the insulating CDW state does not
stay behind in terms of complexity of the electronic structure in comparison
with the metallic regime. The tetramerization of the V ions, structurally an
effective trimerization, leads to quite different behavior. While the (V3,V4)
pair apparently forms a correlated dimer with mixed \A1g/$E_{g1}$ occupation, 
the (V1,V2) ions are strongly orbitally polarized with major $E_{g1}$ occupation and 
negligible coupling. Note however that this picture is of course basis dependent,
and we worked always in the Wannier basis derived from the LDA hamiltonian. Hence
a new orbital basis may be found in the interacting regime where e.g. the occupation 
for (V3,V4) has also polarized character. Concerning the ``free'' spins on (V1,V2),
note that they are still coupled to the neighboring dimers, whereby their degrees of 
freedom are substantially reduced.
The latter effect may serve as an explanation of the quenched local moments
observed below \Tmit.  Fagot {\sl et al.}~\cite{fag06} proposed from anomalous
x-ray measurements a dominant $E_{g1}$ occupation on V1 and an additional dominant
occupation of \A1g on V3, as well as no definite preferential occupation
on (V2,V4). This would describe an orbital order modulated with $2c$ along the
chain. Our picture differs by the fact that we do not find a dominant \A1g
occupation, but rather {\sl two} ions, i.e., (V1,V2), with dominant $E_{g1}$ weight.
Although from a local structural point of view the former proposition appears 
meaningful, the intersite correlation effects appear efficient in singling out
dimer and isolated behavior. Note that dimer formation and nearest-neighbor spins
may also be energetic favorable, since the former brings in some residual hopping
whereas the latter yields energy from spin exchange. Recently, Fazekas 
{\sl et al.}~\cite{faz07} proposed a minimal one-dimensional model, expecting 
thereby also nearest-neighbor $E_g$ spins. However the orbital degeneracy of 
the $E_g$ multiplet, ascribed to spin-orbit coupling, was kept in that model.

The aspect of the spin degree of freedom in \bavs3 was so far only scarcely 
discussed. Albeit it is expected that the former is an important ingredient in 
the understanding of the physics, its role in the different phases is still quite
open. Nakamura and coworkers~\cite{nak94} reported anomalies in the temperature
derivative of the magnetic susceptibility not only at \Tmit but also at the
onset of the bad-metal regime ($\sim$160 K). Hence the detailed role of the local
$E_g$ spins in the metal still poses important questions. Our study of the 
insulator renders it difficult to make precise statements about possible 
long-range order effects for the spin and orbitals, thus leaving questions 
concerning the classification of the electronic phase open. Furthermore, the
mechanisms in conjunction with the famous magnetic transition at $T_{\rm X}$ 
need to be addressed. LDA+(C)DMFT computations may still help in delivering some
further information associated with those problems. For instance, the calculation
of magnetic susceptibilities in the insulating regime, the study of
exchange interactions, as well as the investigation of the influence of the 
interchain coupling especially on the spin arrangement. More sophisticated model
studies for multiorbital chains, perhaps with therefore well-adapted theoretical 
tools like density matrix renormalization group (DMRG)~\cite{whi92}, may also 
deliver important new insights. 

\acknowledgements
We are indebted to A.~Poteryaev, M.~Posternak, A.~Yamasaki and O.K.~Andersen,
as well as S.~Fagot, P.~Foury-Leylekian, J.-P.~Pouget and S.~Ravy for
useful discussions and remarks. This work has been supported by the European 
Union (under contract ``Psi-k f-electrons'' HPRN-CT-2002-00295), the CNRS and 
Ecole Polytechnique. Financial support was provided by the ``Psi-k $f$-electron'' 
Network under contract No. HPRN-CT-2002-00295. Computations were performed at 
IDRIS Orsay.

\bibliographystyle{apsrev}
\bibliography{bibag,bibextra,pubag}

\begin{thebibliography}{41}
\expandafter\ifx\csname natexlab\endcsname\relax\def\natexlab#1{#1}\fi
\expandafter\ifx\csname bibnamefont\endcsname\relax
  \def\bibnamefont#1{#1}\fi
\expandafter\ifx\csname bibfnamefont\endcsname\relax
  \def\bibfnamefont#1{#1}\fi
\expandafter\ifx\csname citenamefont\endcsname\relax
  \def\citenamefont#1{#1}\fi
\expandafter\ifx\csname url\endcsname\relax
  \def\url#1{\texttt{#1}}\fi
\expandafter\ifx\csname urlprefix\endcsname\relax\def\urlprefix{URL }\fi
\providecommand{\bibinfo}[2]{#2}
\providecommand{\eprint}[2][]{\url{#2}}

\bibitem[{\citenamefont{Gardner et~al.}(1969)\citenamefont{Gardner, Vlasse, and
  Wold}}]{gardner_actacry_1969}
\bibinfo{author}{\bibfnamefont{R.}~\bibnamefont{Gardner}},
  \bibinfo{author}{\bibfnamefont{M.}~\bibnamefont{Vlasse}}, \bibnamefont{and}
  \bibinfo{author}{\bibfnamefont{A.}~\bibnamefont{Wold}},
  \bibinfo{journal}{Acta Crystallogr. B} \textbf{\bibinfo{volume}{25}},
  \bibinfo{pages}{781} (\bibinfo{year}{1969}).

\bibitem[{\citenamefont{Massenet et~al.}(1979)\citenamefont{Massenet, Since,
  Mercier, Avignon, Buder, and Nguyen}}]{massenet_jpcsol_1979}
\bibinfo{author}{\bibfnamefont{O.}~\bibnamefont{Massenet}},
  \bibinfo{author}{\bibfnamefont{J.}~\bibnamefont{Since}},
  \bibinfo{author}{\bibfnamefont{J.}~\bibnamefont{Mercier}},
  \bibinfo{author}{\bibfnamefont{M.}~\bibnamefont{Avignon}},
  \bibinfo{author}{\bibfnamefont{R.}~\bibnamefont{Buder}}, \bibnamefont{and}
  \bibinfo{author}{\bibfnamefont{V.}~\bibnamefont{Nguyen}},
  \bibinfo{journal}{J. Phys. Chem. Solids} \textbf{\bibinfo{volume}{40}},
  \bibinfo{pages}{573} (\bibinfo{year}{1979}).

\bibitem[{\citenamefont{Matsuhara et~al.}(1991)\citenamefont{Matsuhara, Wada,
  Nakamizo, Yamauchi, and Tanaka}}]{mat91}
\bibinfo{author}{\bibfnamefont{K.}~\bibnamefont{Matsuhara}},
  \bibinfo{author}{\bibfnamefont{T.}~\bibnamefont{Wada}},
  \bibinfo{author}{\bibfnamefont{T.}~\bibnamefont{Nakamizo}},
  \bibinfo{author}{\bibfnamefont{H.}~\bibnamefont{Yamauchi}}, \bibnamefont{and}
  \bibinfo{author}{\bibfnamefont{S.}~\bibnamefont{Tanaka}},
  \bibinfo{journal}{Phys. Rev. B} \textbf{\bibinfo{volume}{43}},
  \bibinfo{pages}{13118} (\bibinfo{year}{1991}).

\bibitem[{\citenamefont{{Graf} et~al.}(1995)\citenamefont{{Graf}, {Mandrus},
  {Lawrence}, {Thompson}, {Canfield}, {Cheong}, and
  {Rupp}}}]{graf_bavs3_pressure_prb_1995}
\bibinfo{author}{\bibfnamefont{T.}~\bibnamefont{{Graf}}},
  \bibinfo{author}{\bibfnamefont{D.}~\bibnamefont{{Mandrus}}},
  \bibinfo{author}{\bibfnamefont{J.~M.} \bibnamefont{{Lawrence}}},
  \bibinfo{author}{\bibfnamefont{J.~D.} \bibnamefont{{Thompson}}},
  \bibinfo{author}{\bibfnamefont{P.~C.} \bibnamefont{{Canfield}}},
  \bibinfo{author}{\bibfnamefont{S.-W.} \bibnamefont{{Cheong}}},
  \bibnamefont{and} \bibinfo{author}{\bibfnamefont{L.~W.}
  \bibnamefont{{Rupp}}}, \bibinfo{journal}{Phys. Rev. B}
  \textbf{\bibinfo{volume}{51}}, \bibinfo{pages}{2037} (\bibinfo{year}{1995}).

\bibitem[{\citenamefont{{Booth} et~al.}(1999)\citenamefont{{Booth}, {Figueroa},
  {Lawrence}, {Hundley}, and {Thompson}}}]{boo99}
\bibinfo{author}{\bibfnamefont{C.~H.} \bibnamefont{{Booth}}},
  \bibinfo{author}{\bibfnamefont{E.}~\bibnamefont{{Figueroa}}},
  \bibinfo{author}{\bibfnamefont{J.~M.} \bibnamefont{{Lawrence}}},
  \bibinfo{author}{\bibfnamefont{M.~F.} \bibnamefont{{Hundley}}},
  \bibnamefont{and} \bibinfo{author}{\bibfnamefont{J.~D.}
  \bibnamefont{{Thompson}}}, \bibinfo{journal}{Phys. Rev. B}
  \textbf{\bibinfo{volume}{60}}, \bibinfo{pages}{14852} (\bibinfo{year}{1999}).

\bibitem[{\citenamefont{{Whangbo} et~al.}(2003)\citenamefont{{Whangbo}, {Koo},
  {Dai}, and {Villesuzanne}}}]{whangbo_bavs3_puzzling_jsschem_2003}
\bibinfo{author}{\bibfnamefont{M.-H.} \bibnamefont{{Whangbo}}},
  \bibinfo{author}{\bibfnamefont{H.-J.} \bibnamefont{{Koo}}},
  \bibinfo{author}{\bibfnamefont{D.}~\bibnamefont{{Dai}}}, \bibnamefont{and}
  \bibinfo{author}{\bibfnamefont{A.}~\bibnamefont{{Villesuzanne}}},
  \bibinfo{journal}{J. Solid State Chem.} \textbf{\bibinfo{volume}{175}},
  \bibinfo{pages}{384} (\bibinfo{year}{2003}).

\bibitem[{\citenamefont{{Inami} et~al.}(2002)\citenamefont{{Inami}, {Ohwada},
  {Kimura}, {Watanabe}, {Noda}, {Nakamura}, {Yamasaki}, {Shiga}, {Ikeda}, and
  {Murakami}}}]{inami_bavs3_prb_2002}
\bibinfo{author}{\bibfnamefont{T.}~\bibnamefont{{Inami}}},
  \bibinfo{author}{\bibfnamefont{K.}~\bibnamefont{{Ohwada}}},
  \bibinfo{author}{\bibfnamefont{H.}~\bibnamefont{{Kimura}}},
  \bibinfo{author}{\bibfnamefont{M.}~\bibnamefont{{Watanabe}}},
  \bibinfo{author}{\bibfnamefont{Y.}~\bibnamefont{{Noda}}},
  \bibinfo{author}{\bibfnamefont{H.}~\bibnamefont{{Nakamura}}},
  \bibinfo{author}{\bibfnamefont{T.}~\bibnamefont{{Yamasaki}}},
  \bibinfo{author}{\bibfnamefont{M.}~\bibnamefont{{Shiga}}},
  \bibinfo{author}{\bibfnamefont{N.}~\bibnamefont{{Ikeda}}}, \bibnamefont{and}
  \bibinfo{author}{\bibfnamefont{Y.}~\bibnamefont{{Murakami}}},
  \bibinfo{journal}{Phys. Rev. B} \textbf{\bibinfo{volume}{66}},
  \bibinfo{pages}{073108} (\bibinfo{year}{2002}).

\bibitem[{\citenamefont{Fagot et~al.}(2005)\citenamefont{Fagot,
  Foury-Leylekian, Ravy, Pouget, Anne, Popov, Lobanov, and Greenblatt}}]{fag05}
\bibinfo{author}{\bibfnamefont{S.}~\bibnamefont{Fagot}},
  \bibinfo{author}{\bibfnamefont{P.}~\bibnamefont{Foury-Leylekian}},
  \bibinfo{author}{\bibfnamefont{S.}~\bibnamefont{Ravy}},
  \bibinfo{author}{\bibfnamefont{J.-P.} \bibnamefont{Pouget}},
  \bibinfo{author}{\bibfnamefont{M.}~\bibnamefont{Anne}},
  \bibinfo{author}{\bibfnamefont{G.}~\bibnamefont{Popov}},
  \bibinfo{author}{\bibfnamefont{M.~V.} \bibnamefont{Lobanov}},
  \bibnamefont{and}
  \bibinfo{author}{\bibfnamefont{M.}~\bibnamefont{Greenblatt}},
  \bibinfo{journal}{Solid State Sciences} \textbf{\bibinfo{volume}{7}},
  \bibinfo{pages}{718} (\bibinfo{year}{2005}).

\bibitem[{\citenamefont{Nakamura et~al.}(2000)\citenamefont{Nakamura, Yamasaki,
  Giri, Imai, Shiga, Kojima, Nishi, and Metoki}}]{nak00}
\bibinfo{author}{\bibfnamefont{H.}~\bibnamefont{Nakamura}},
  \bibinfo{author}{\bibfnamefont{T.}~\bibnamefont{Yamasaki}},
  \bibinfo{author}{\bibfnamefont{S.}~\bibnamefont{Giri}},
  \bibinfo{author}{\bibfnamefont{H.}~\bibnamefont{Imai}},
  \bibinfo{author}{\bibfnamefont{M.}~\bibnamefont{Shiga}},
  \bibinfo{author}{\bibfnamefont{K.}~\bibnamefont{Kojima}},
  \bibinfo{author}{\bibfnamefont{M.}~\bibnamefont{Nishi}}, \bibnamefont{and}
  \bibinfo{author}{\bibfnamefont{K.~K.~N.} \bibnamefont{Metoki}},
  \bibinfo{journal}{J. Phys. Soc. Jpn.} \textbf{\bibinfo{volume}{69}},
  \bibinfo{pages}{2763} (\bibinfo{year}{2000}).

\bibitem[{\citenamefont{Higemoto et~al.}(202)\citenamefont{Higemoto, Koda,
  Maruta, Nishiyama, Nakamura, Giri, and Shiga}}]{hig02}
\bibinfo{author}{\bibfnamefont{W.}~\bibnamefont{Higemoto}},
  \bibinfo{author}{\bibfnamefont{A.}~\bibnamefont{Koda}},
  \bibinfo{author}{\bibfnamefont{G.}~\bibnamefont{Maruta}},
  \bibinfo{author}{\bibfnamefont{K.}~\bibnamefont{Nishiyama}},
  \bibinfo{author}{\bibfnamefont{H.}~\bibnamefont{Nakamura}},
  \bibinfo{author}{\bibfnamefont{S.}~\bibnamefont{Giri}}, \bibnamefont{and}
  \bibinfo{author}{\bibfnamefont{M.}~\bibnamefont{Shiga}}, \bibinfo{journal}{J.
  Phys. Soc. Jpn.} \textbf{\bibinfo{volume}{71}}, \bibinfo{pages}{2361}
  (\bibinfo{year}{202}).

\bibitem[{\citenamefont{{Forr{\' o}} et~al.}(2000)\citenamefont{{Forr{\' o}},
  {Ga{\' a}l}, {Berger}, {Fazekas}, {Penc}, {K{\' e}zsm{\' a}rki}, and {Mih{\'
  a}ly}}}]{forro_bavs3_qcp_prl_2000}
\bibinfo{author}{\bibfnamefont{L.}~\bibnamefont{{Forr{\' o}}}},
  \bibinfo{author}{\bibfnamefont{R.}~\bibnamefont{{Ga{\' a}l}}},
  \bibinfo{author}{\bibfnamefont{H.}~\bibnamefont{{Berger}}},
  \bibinfo{author}{\bibfnamefont{P.}~\bibnamefont{{Fazekas}}},
  \bibinfo{author}{\bibfnamefont{K.}~\bibnamefont{{Penc}}},
  \bibinfo{author}{\bibfnamefont{I.}~\bibnamefont{{K{\' e}zsm{\' a}rki}}},
  \bibnamefont{and} \bibinfo{author}{\bibfnamefont{G.}~\bibnamefont{{Mih{\'
  a}ly}}}, \bibinfo{journal}{Phys. Rev. Lett.} \textbf{\bibinfo{volume}{85}},
  \bibinfo{pages}{1938} (\bibinfo{year}{2000}).

\bibitem[{\citenamefont{Barisi{\'c} et~al.}(2006)\citenamefont{Barisi{\'c},
  K{\' e}zsm{\' a}rki, Fazekas, , Mih{\'a}ly, Berger, Demk{\'o}, and
  Forr{\'o}}}]{Bar06}
\bibinfo{author}{\bibfnamefont{N.}~\bibnamefont{Barisi{\'c}}},
  \bibinfo{author}{\bibfnamefont{I.}~\bibnamefont{K{\' e}zsm{\' a}rki}},
  \bibinfo{author}{\bibfnamefont{P.}~\bibnamefont{Fazekas}}, ,
  \bibinfo{author}{\bibfnamefont{G.}~\bibnamefont{Mih{\'a}ly}},
  \bibinfo{author}{\bibfnamefont{H.}~\bibnamefont{Berger}},
  \bibinfo{author}{\bibfnamefont{L.}~\bibnamefont{Demk{\'o}}},
  \bibnamefont{and}
  \bibinfo{author}{\bibfnamefont{L.}~\bibnamefont{Forr{\'o}}},
  \bibinfo{journal}{cond-mat/0602262}  (\bibinfo{year}{2006}).

\bibitem[{\citenamefont{{Fagot} et~al.}(2003)\citenamefont{{Fagot},
  {Foury-Leylekian}, {Ravy}, {Pouget}, and {Berger}}}]{fagot_bavs3_prl_2003}
\bibinfo{author}{\bibfnamefont{S.}~\bibnamefont{{Fagot}}},
  \bibinfo{author}{\bibfnamefont{P.}~\bibnamefont{{Foury-Leylekian}}},
  \bibinfo{author}{\bibfnamefont{S.}~\bibnamefont{{Ravy}}},
  \bibinfo{author}{\bibfnamefont{J.}~\bibnamefont{{Pouget}}}, \bibnamefont{and}
  \bibinfo{author}{\bibfnamefont{H.}~\bibnamefont{{Berger}}},
  \bibinfo{journal}{Phys. Rev. Lett.} \textbf{\bibinfo{volume}{90}},
  \bibinfo{pages}{196401} (\bibinfo{year}{2003}).

\bibitem[{\citenamefont{Lechermann et~al.}(2006)\citenamefont{Lechermann,
  Georges, Poteryaev, Biermann, Posternak, Yamasaki, and Andersen}}]{lec06}
\bibinfo{author}{\bibfnamefont{F.}~\bibnamefont{Lechermann}},
  \bibinfo{author}{\bibfnamefont{A.}~\bibnamefont{Georges}},
  \bibinfo{author}{\bibfnamefont{A.}~\bibnamefont{Poteryaev}},
  \bibinfo{author}{\bibfnamefont{S.}~\bibnamefont{Biermann}},
  \bibinfo{author}{\bibfnamefont{M.}~\bibnamefont{Posternak}},
  \bibinfo{author}{\bibfnamefont{A.}~\bibnamefont{Yamasaki}}, \bibnamefont{and}
  \bibinfo{author}{\bibfnamefont{O.}~\bibnamefont{Andersen}},
  \bibinfo{journal}{Phys. Rev. B} \textbf{\bibinfo{volume}{74}},
  \bibinfo{pages}{125120} (\bibinfo{year}{2006}).

\bibitem[{\citenamefont{Mih{\'a}ly et~al.}(2000)\citenamefont{Mih{\'a}ly,
  K{\'e}zsm{\'a}rki, Z{\'a}mborszky, Miljak, Penc, Fazekas, Berger, and
  Forr{\'o}}}]{mih00}
\bibinfo{author}{\bibfnamefont{G.}~\bibnamefont{Mih{\'a}ly}},
  \bibinfo{author}{\bibfnamefont{I.}~\bibnamefont{K{\'e}zsm{\'a}rki}},
  \bibinfo{author}{\bibfnamefont{F.}~\bibnamefont{Z{\'a}mborszky}},
  \bibinfo{author}{\bibfnamefont{M.}~\bibnamefont{Miljak}},
  \bibinfo{author}{\bibfnamefont{K.}~\bibnamefont{Penc}},
  \bibinfo{author}{\bibfnamefont{P.}~\bibnamefont{Fazekas}},
  \bibinfo{author}{\bibfnamefont{H.}~\bibnamefont{Berger}}, \bibnamefont{and}
  \bibinfo{author}{\bibfnamefont{L.}~\bibnamefont{Forr{\'o}}},
  \bibinfo{journal}{Phys. Rev. B} \textbf{\bibinfo{volume}{61}},
  \bibinfo{pages}{R7831} (\bibinfo{year}{2000}).

\bibitem[{\citenamefont{Imai et~al.}(1996)\citenamefont{Imai, Wada, and
  Shiga}}]{ima96}
\bibinfo{author}{\bibfnamefont{H.}~\bibnamefont{Imai}},
  \bibinfo{author}{\bibfnamefont{H.}~\bibnamefont{Wada}}, \bibnamefont{and}
  \bibinfo{author}{\bibfnamefont{M.}~\bibnamefont{Shiga}}, \bibinfo{journal}{J.
  Phys. Soc. Jpn.} \textbf{\bibinfo{volume}{65}}, \bibinfo{pages}{3460}
  (\bibinfo{year}{1996}).

\bibitem[{\citenamefont{Fagot et~al.}(2006)\citenamefont{Fagot,
  Foury-Leylekian, Ravy, Pouget, Lorenzo, Joly, Greenblatt, Lobanov, and
  Popov}}]{fag06}
\bibinfo{author}{\bibfnamefont{S.}~\bibnamefont{Fagot}},
  \bibinfo{author}{\bibfnamefont{P.}~\bibnamefont{Foury-Leylekian}},
  \bibinfo{author}{\bibfnamefont{S.}~\bibnamefont{Ravy}},
  \bibinfo{author}{\bibfnamefont{J.-P.} \bibnamefont{Pouget}},
  \bibinfo{author}{\bibfnamefont{E.}~\bibnamefont{Lorenzo}},
  \bibinfo{author}{\bibfnamefont{Y.}~\bibnamefont{Joly}},
  \bibinfo{author}{\bibfnamefont{M.}~\bibnamefont{Greenblatt}},
  \bibinfo{author}{\bibfnamefont{M.~V.} \bibnamefont{Lobanov}},
  \bibnamefont{and} \bibinfo{author}{\bibfnamefont{G.}~\bibnamefont{Popov}},
  \bibinfo{journal}{Phys. Rev. B} \textbf{\bibinfo{volume}{73}},
  \bibinfo{pages}{033102} (\bibinfo{year}{2006}).

\bibitem[{\citenamefont{Nakamura et~al.}(1994)\citenamefont{Nakamura, Sekiyama,
  Namatame, Fujimori, Yoshihara, Ohtani, Misu, and Takano}}]{nak94}
\bibinfo{author}{\bibfnamefont{M.}~\bibnamefont{Nakamura}},
  \bibinfo{author}{\bibfnamefont{A.}~\bibnamefont{Sekiyama}},
  \bibinfo{author}{\bibfnamefont{H.}~\bibnamefont{Namatame}},
  \bibinfo{author}{\bibfnamefont{A.}~\bibnamefont{Fujimori}},
  \bibinfo{author}{\bibfnamefont{H.}~\bibnamefont{Yoshihara}},
  \bibinfo{author}{\bibfnamefont{T.}~\bibnamefont{Ohtani}},
  \bibinfo{author}{\bibfnamefont{A.}~\bibnamefont{Misu}}, \bibnamefont{and}
  \bibinfo{author}{\bibfnamefont{M.}~\bibnamefont{Takano}},
  \bibinfo{journal}{Phys. Rev. B} \textbf{\bibinfo{volume}{49}},
  \bibinfo{pages}{16191} (\bibinfo{year}{1994}).

\bibitem[{\citenamefont{K{\'e}zsm{\'a}rki
  et~al.}(2006)\citenamefont{K{\'e}zsm{\'a}rki, Mih{\'a}ly, Ga{\'a}l,
  Bari\v{s}i\'{c}, Akrap, Berger, Forr{\'o}, Homes, and Mih{\'a}ly}}]{kez06}
\bibinfo{author}{\bibfnamefont{I.}~\bibnamefont{K{\'e}zsm{\'a}rki}},
  \bibinfo{author}{\bibfnamefont{G.}~\bibnamefont{Mih{\'a}ly}},
  \bibinfo{author}{\bibfnamefont{R.}~\bibnamefont{Ga{\'a}l}},
  \bibinfo{author}{\bibfnamefont{N.}~\bibnamefont{Bari\v{s}i\'{c}}},
  \bibinfo{author}{\bibfnamefont{A.}~\bibnamefont{Akrap}},
  \bibinfo{author}{\bibfnamefont{H.}~\bibnamefont{Berger}},
  \bibinfo{author}{\bibfnamefont{L.}~\bibnamefont{Forr{\'o}}},
  \bibinfo{author}{\bibfnamefont{C.~C.} \bibnamefont{Homes}}, \bibnamefont{and}
  \bibinfo{author}{\bibfnamefont{L.}~\bibnamefont{Mih{\'a}ly}},
  \bibinfo{journal}{Phys. Rev. Lett.} \textbf{\bibinfo{volume}{96}},
  \bibinfo{pages}{186402} (\bibinfo{year}{2006}).

\bibitem[{\citenamefont{Nakamura et~al.}(1997)\citenamefont{Nakamura, Imai, and
  Shiga}}]{nak97}
\bibinfo{author}{\bibfnamefont{M.}~\bibnamefont{Nakamura}},
  \bibinfo{author}{\bibfnamefont{H.}~\bibnamefont{Imai}}, \bibnamefont{and}
  \bibinfo{author}{\bibfnamefont{M.}~\bibnamefont{Shiga}},
  \bibinfo{journal}{Phys. Rev. Lett.} \textbf{\bibinfo{volume}{79}},
  \bibinfo{pages}{3779} (\bibinfo{year}{1997}).

\bibitem[{\citenamefont{M{\'a}lek et~al.}(2003)\citenamefont{M{\'a}lek,
  Drechsler, Flach, Jeckelmann, and Kladko}}]{mal03}
\bibinfo{author}{\bibfnamefont{J.}~\bibnamefont{M{\'a}lek}},
  \bibinfo{author}{\bibfnamefont{S.-L.} \bibnamefont{Drechsler}},
  \bibinfo{author}{\bibfnamefont{S.}~\bibnamefont{Flach}},
  \bibinfo{author}{\bibfnamefont{E.}~\bibnamefont{Jeckelmann}},
  \bibnamefont{and} \bibinfo{author}{\bibfnamefont{K.}~\bibnamefont{Kladko}},
  \bibinfo{journal}{J. Phys. Soc. Jpn.} \textbf{\bibinfo{volume}{72}},
  \bibinfo{pages}{2277} (\bibinfo{year}{2003}).

\bibitem[{\citenamefont{Lechermann et~al.}(2005)\citenamefont{Lechermann,
  Biermann, and Georges}}]{lec05}
\bibinfo{author}{\bibfnamefont{F.}~\bibnamefont{Lechermann}},
  \bibinfo{author}{\bibfnamefont{S.}~\bibnamefont{Biermann}}, \bibnamefont{and}
  \bibinfo{author}{\bibfnamefont{A.}~\bibnamefont{Georges}},
  \bibinfo{journal}{Phys. Rev. Lett.} \textbf{\bibinfo{volume}{94}},
  \bibinfo{pages}{166402} (\bibinfo{year}{2005}).

\bibitem[{\citenamefont{{Anisimov} et~al.}(1997)\citenamefont{{Anisimov},
  {Poteryaev}, {Korotin}, {Anokhin}, and {Kotliar}}}]{anisimov_lda+dmft_1997}
\bibinfo{author}{\bibfnamefont{V.~I.} \bibnamefont{{Anisimov}}},
  \bibinfo{author}{\bibfnamefont{A.~I.} \bibnamefont{{Poteryaev}}},
  \bibinfo{author}{\bibfnamefont{M.~A.} \bibnamefont{{Korotin}}},
  \bibinfo{author}{\bibfnamefont{A.~O.} \bibnamefont{{Anokhin}}},
  \bibnamefont{and}
  \bibinfo{author}{\bibfnamefont{G.}~\bibnamefont{{Kotliar}}},
  \bibinfo{journal}{J. Phys. Cond. Matter} \textbf{\bibinfo{volume}{9}},
  \bibinfo{pages}{7359} (\bibinfo{year}{1997}).

\bibitem[{\citenamefont{{Lichtenstein} and
  {Katsnelson}}(1998)}]{lichtenstein_lda+dmft_1998}
\bibinfo{author}{\bibfnamefont{A.~I.} \bibnamefont{{Lichtenstein}}}
  \bibnamefont{and} \bibinfo{author}{\bibfnamefont{M.~I.}
  \bibnamefont{{Katsnelson}}}, \bibinfo{journal}{Phys. Rev. B}
  \textbf{\bibinfo{volume}{57}}, \bibinfo{pages}{6884} (\bibinfo{year}{1998}).

\bibitem[{\citenamefont{Meyer et~al.}(unpublished)\citenamefont{Meyer,
  Els\"{a}sser, Lechermann, and F\"{a}hnle}}]{mbpp_code}
\bibinfo{author}{\bibfnamefont{B.}~\bibnamefont{Meyer}},
  \bibinfo{author}{\bibfnamefont{C.}~\bibnamefont{Els\"{a}sser}},
  \bibinfo{author}{\bibfnamefont{F.}~\bibnamefont{Lechermann}},
  \bibnamefont{and}
  \bibinfo{author}{\bibfnamefont{M.}~\bibnamefont{F\"{a}hnle}},
  \emph{\bibinfo{title}{FORTRAN 90 Program for Mixed-Basis-Pseudopotential
  Calculations for Crystals}}, \bibinfo{organization}{Max-Planck-Institut
  f\"{u}r Metallforschung, Stuttgart} (\bibinfo{year}{unpublished}).

\bibitem[{\citenamefont{Marzari and Vanderbilt}(1997)}]{mar97}
\bibinfo{author}{\bibfnamefont{N.}~\bibnamefont{Marzari}} \bibnamefont{and}
  \bibinfo{author}{\bibfnamefont{D.}~\bibnamefont{Vanderbilt}},
  \bibinfo{journal}{Phys. Rev. B} \textbf{\bibinfo{volume}{56}},
  \bibinfo{pages}{12847} (\bibinfo{year}{1997}).

\bibitem[{\citenamefont{Souza et~al.}(2001)\citenamefont{Souza, Marzari, and
  Vanderbilt}}]{sou01}
\bibinfo{author}{\bibfnamefont{I.}~\bibnamefont{Souza}},
  \bibinfo{author}{\bibfnamefont{N.}~\bibnamefont{Marzari}}, \bibnamefont{and}
  \bibinfo{author}{\bibfnamefont{D.}~\bibnamefont{Vanderbilt}},
  \bibinfo{journal}{Phys. Rev. B} \textbf{\bibinfo{volume}{65}},
  \bibinfo{pages}{035109} (\bibinfo{year}{2001}).

\bibitem[{\citenamefont{Castellani et~al.}(1978)\citenamefont{Castellani,
  Natoli, and Ranninger}}]{cas78}
\bibinfo{author}{\bibfnamefont{C.}~\bibnamefont{Castellani}},
  \bibinfo{author}{\bibfnamefont{C.~R.} \bibnamefont{Natoli}},
  \bibnamefont{and}
  \bibinfo{author}{\bibfnamefont{J.}~\bibnamefont{Ranninger}},
  \bibinfo{journal}{Phys. Rev. B} \textbf{\bibinfo{volume}{18}},
  \bibinfo{pages}{4945} (\bibinfo{year}{1978}).

\bibitem[{\citenamefont{Fr\'{e}sard and Kotliar}(1997)}]{fre97}
\bibinfo{author}{\bibfnamefont{R.}~\bibnamefont{Fr\'{e}sard}} \bibnamefont{and}
  \bibinfo{author}{\bibfnamefont{G.}~\bibnamefont{Kotliar}},
  \bibinfo{journal}{Phys. Rev. B} \textbf{\bibinfo{volume}{56}},
  \bibinfo{pages}{12909} (\bibinfo{year}{1997}).

\bibitem[{\citenamefont{Hirsch and Fye}(1986)}]{hirsch_fye}
\bibinfo{author}{\bibfnamefont{J.~E.} \bibnamefont{Hirsch}} \bibnamefont{and}
  \bibinfo{author}{\bibfnamefont{R.~M.} \bibnamefont{Fye}},
  \bibinfo{journal}{Phys. Rev. Lett.} \textbf{\bibinfo{volume}{25}},
  \bibinfo{pages}{2521} (\bibinfo{year}{1986}).

\bibitem[{\citenamefont{Biroli et~al.}(2004)\citenamefont{Biroli, Parcollet,
  and Kotliar}}]{bir04}
\bibinfo{author}{\bibfnamefont{G.}~\bibnamefont{Biroli}},
  \bibinfo{author}{\bibfnamefont{O.}~\bibnamefont{Parcollet}},
  \bibnamefont{and} \bibinfo{author}{\bibfnamefont{G.}~\bibnamefont{Kotliar}},
  \bibinfo{journal}{Phys. Rev. B} \textbf{\bibinfo{volume}{69}},
  \bibinfo{pages}{205108} (\bibinfo{year}{2004}).

\bibitem[{\citenamefont{Maier et~al.}(2005)\citenamefont{Maier, Jarrell,
  Pruschke, and Hettler}}]{mai05}
\bibinfo{author}{\bibfnamefont{T.}~\bibnamefont{Maier}},
  \bibinfo{author}{\bibfnamefont{M.}~\bibnamefont{Jarrell}},
  \bibinfo{author}{\bibfnamefont{T.}~\bibnamefont{Pruschke}}, \bibnamefont{and}
  \bibinfo{author}{\bibfnamefont{M.~H.} \bibnamefont{Hettler}},
  \bibinfo{journal}{Rev. Mod. Phys.} \textbf{\bibinfo{volume}{77}},
  \bibinfo{pages}{1027} (\bibinfo{year}{2005}).

\bibitem[{\citenamefont{Lichtenstein et~al.}(2002)\citenamefont{Lichtenstein,
  Katsnelson, and Kotliar}}]{lic02}
\bibinfo{author}{\bibfnamefont{A.}~\bibnamefont{Lichtenstein}},
  \bibinfo{author}{\bibfnamefont{M.}~\bibnamefont{Katsnelson}},
  \bibnamefont{and} \bibinfo{author}{\bibfnamefont{G.}~\bibnamefont{Kotliar}},
  \bibinfo{journal}{cond-mat/0211076}  (\bibinfo{year}{2002}).

\bibitem[{\citenamefont{Poteryaev et~al.}(2004)\citenamefont{Poteryaev,
  Lichtenstein, and Kotliar}}]{pot04}
\bibinfo{author}{\bibfnamefont{A.~I.} \bibnamefont{Poteryaev}},
  \bibinfo{author}{\bibfnamefont{A.~I.} \bibnamefont{Lichtenstein}},
  \bibnamefont{and} \bibinfo{author}{\bibfnamefont{G.}~\bibnamefont{Kotliar}},
  \bibinfo{journal}{Phys. Rev. Lett.} \textbf{\bibinfo{volume}{93}},
  \bibinfo{pages}{086401} (\bibinfo{year}{2004}).

\bibitem[{\citenamefont{Biermann et~al.}(2005)\citenamefont{Biermann,
  Poteryaev, Lichtenstein, and Georges}}]{bie05}
\bibinfo{author}{\bibfnamefont{S.}~\bibnamefont{Biermann}},
  \bibinfo{author}{\bibfnamefont{A.}~\bibnamefont{Poteryaev}},
  \bibinfo{author}{\bibfnamefont{A.~I.} \bibnamefont{Lichtenstein}},
  \bibnamefont{and} \bibinfo{author}{\bibfnamefont{A.}~\bibnamefont{Georges}},
  \bibinfo{journal}{Phys. Rev. Lett.} \textbf{\bibinfo{volume}{94}},
  \bibinfo{pages}{026404} (\bibinfo{year}{2005}).

\bibitem[{\citenamefont{{Ghedira} et~al.}(1986)\citenamefont{{Ghedira}, {Anne},
  {Chenavas}, {Marezio}, and {Sayetat}}}]{ghedira_bavs3_neutrons_jpc_1986}
\bibinfo{author}{\bibfnamefont{M.}~\bibnamefont{{Ghedira}}},
  \bibinfo{author}{\bibfnamefont{M.}~\bibnamefont{{Anne}}},
  \bibinfo{author}{\bibfnamefont{J.}~\bibnamefont{{Chenavas}}},
  \bibinfo{author}{\bibfnamefont{M.}~\bibnamefont{{Marezio}}},
  \bibnamefont{and}
  \bibinfo{author}{\bibfnamefont{F.}~\bibnamefont{{Sayetat}}},
  \bibinfo{journal}{Journal of Physics C Solid State Physics}
  \textbf{\bibinfo{volume}{19}}, \bibinfo{pages}{6489} (\bibinfo{year}{1986}).

\bibitem[{\citenamefont{Mattheiss}(1995)}]{mattheiss_bavs3_1995}
\bibinfo{author}{\bibfnamefont{L.}~\bibnamefont{Mattheiss}},
  \bibinfo{journal}{Solid State Commun.} \textbf{\bibinfo{volume}{93}},
  \bibinfo{pages}{791} (\bibinfo{year}{1995}).

\bibitem[{\citenamefont{{Whangbo} et~al.}(2002)\citenamefont{{Whangbo}, {Koo},
  {Dai}, and {Villesuzanne}}}]{whangbo_bavs3_jsschem_2002}
\bibinfo{author}{\bibfnamefont{M.-H.} \bibnamefont{{Whangbo}}},
  \bibinfo{author}{\bibfnamefont{H.-J.} \bibnamefont{{Koo}}},
  \bibinfo{author}{\bibfnamefont{D.}~\bibnamefont{{Dai}}}, \bibnamefont{and}
  \bibinfo{author}{\bibfnamefont{A.}~\bibnamefont{{Villesuzanne}}},
  \bibinfo{journal}{J. Solid State Chem.} \textbf{\bibinfo{volume}{165}},
  \bibinfo{pages}{345} (\bibinfo{year}{2002}).

\bibitem[{\citenamefont{Biermann et~al.}(2001)\citenamefont{Biermann, Georges,
  Lichtenstein, and Giamarchi}}]{bie01}
\bibinfo{author}{\bibfnamefont{S.}~\bibnamefont{Biermann}},
  \bibinfo{author}{\bibfnamefont{A.}~\bibnamefont{Georges}},
  \bibinfo{author}{\bibfnamefont{A.}~\bibnamefont{Lichtenstein}},
  \bibnamefont{and}
  \bibinfo{author}{\bibfnamefont{T.}~\bibnamefont{Giamarchi}},
  \bibinfo{journal}{Phys. Rev. Lett.} \textbf{\bibinfo{volume}{87}},
  \bibinfo{pages}{276405} (\bibinfo{year}{2001}).

\bibitem[{\citenamefont{Fazekas et~al.}(2007)\citenamefont{Fazekas, Penc,
  Radn\'{o}czi, Barisi{\'c}, Berger, Forr\'{o}, Mitrovi\'{c}, Gauzzi,
  Demk\'{o}, K\'{e}zsm\'{a}rki et~al.}}]{faz07}
\bibinfo{author}{\bibfnamefont{P.}~\bibnamefont{Fazekas}},
  \bibinfo{author}{\bibfnamefont{K.}~\bibnamefont{Penc}},
  \bibinfo{author}{\bibfnamefont{K.}~\bibnamefont{Radn\'{o}czi}},
  \bibinfo{author}{\bibfnamefont{N.}~\bibnamefont{Barisi{\'c}}},
  \bibinfo{author}{\bibfnamefont{H.}~\bibnamefont{Berger}},
  \bibinfo{author}{\bibfnamefont{L.}~\bibnamefont{Forr\'{o}}},
  \bibinfo{author}{\bibfnamefont{S.}~\bibnamefont{Mitrovi\'{c}}},
  \bibinfo{author}{\bibfnamefont{A.}~\bibnamefont{Gauzzi}},
  \bibinfo{author}{\bibfnamefont{L.}~\bibnamefont{Demk\'{o}}},
  \bibinfo{author}{\bibfnamefont{I.}~\bibnamefont{K\'{e}zsm\'{a}rki}},
  \bibnamefont{et~al.}, \bibinfo{journal}{cond-mat/0702510}
  (\bibinfo{year}{2007}).

\bibitem[{\citenamefont{White}(1992)}]{whi92}
\bibinfo{author}{\bibfnamefont{S.}~\bibnamefont{White}},
  \bibinfo{journal}{Phys. Rev. Lett.} \textbf{\bibinfo{volume}{69}},
  \bibinfo{pages}{2863} (\bibinfo{year}{1992}).

\end{thebibliography}

\end{document}